\documentclass[a4paper,10pt]{article}

\usepackage[T1]{fontenc}

\usepackage[dvips,colorlinks,bookmarksopen,bookmarksnumbered,citecolor=red,urlcolor=red,breaklinks = true]{hyperref}
\usepackage{subfigure,times,epsfig,epsf,latexsym,graphicx,multirow,fullpage}

\usepackage{lscape}
\usepackage[ruled,vlined]{algorithm2e}

\usepackage{amssymb,amsmath,amsthm}
\usepackage{algpseudocode}

\usepackage{xr}
\newtheorem{theorem}{Theorem}
\newtheorem{lemma}{Lemma}
\newtheorem{corollary}{Corollary}
\newtheorem{remark}{Remark}

\makeatletter
\gdef\thetable{A-\@arabic\c@table}
\makeatother

\begin{document}

\title{Multimedia Channel Allocation in Cognitive Radio Networks using FDM-FDMA and OFDM-FDMA \footnote{A preliminary version of this paper appeared in Proc. of \textsl{$3^{rd}$ International Conf. on Communication Systems and Networks (COMSNETS)}, Bangalore, India, Jan. 4-8, 2011, doi: 10.1109/COMSNETS.2011.5716435, under the title "Multimedia Communication in Cognitive Radio Networks based on Sample Division Multiplexing".} \footnote{A version of this paper to be appeared in ``IEEE Transactions on Cognitive Communications and Networking'', under the title "Non-Contiguous Channel Allocation for Multimedia Communication in Cognitive Radio Networks".}}

\author{Ansuman~Bhattacharya\footnote{A. Bhattacharya is with the Dept. of C.S.E., N.I.T., Meghalaya, India e-mail: ansuman@nitm.ac.in},
Rabindranath~Ghosh\footnote{R. Ghosh is with the Dept. of E.C.E., St. Thomas' College of Engg. and Tech., Kolkata, India e-mail: rnghosh@gmail.com},
Koushik~Sinha\footnote{K. Sinha is with the Dept. of C.S., S.I.U., Carbondale, IL, USA e-mail: koushik.sinha@cs.siu.edu}, \\
Debasish~Datta\footnote{D. Datta is with the Dept. of E.E.C.E., I.I.T., Kharagpur, India e-mail: ddatta@ece.iitkgp.ernet.in},
and~Bhabani~P.~Sinha\footnote{B. P. Sinha is with the A.C.M. Unit, I.S.I., Kolkata, India e-mail: bhabani@isical.ac.in}}

\maketitle

\begin{abstract}
In conventional wireless systems, unless a contiguous frequency band with width at least equal to the required bandwidth is obtained, multimedia communication can not be effected with the desired Quality of Service. We propose here a novel channel allocation technique to overcome this limitation in a Cognitive Radio Network which is based on utilizing several non-contiguous channels, each of width smaller than the required bandwidth, but whose sum equals at least the required bandwidth. We present algorithms for channel sensing, channel reservation and channel deallocation along with transmission and reception protocols with two different implementations based on $FDM-FDMA$ and $OFDM-FDMA$ techniques. Simulation results for both these implementations show that the proposed technique outperforms the existing first-fit and best-fit~\cite{b109, b110} allocation techniques in terms of the average number of attempts needed for acquiring the necessary number of channels for all traffic situations ranging from light to extremely heavy traffic. Further, the proposed technique can allocate the required numbers of channels in less than one second with $FDM-FDMA$ ($4.5$ second with $OFDM-FDMA$) even for $96\%$ traffic load, while the first-fit and best-fit techniques fail to allocate any channel in such situations.
\end{abstract}


\section{Introduction}\label{intro}

The concept of {\em Cognitive Radio} ($CR$)~\cite{b31} is based on dividing the available radio spectrum into several parts, with some part reserved for the licensed users and the rest freely available for all. A {\em Cognitive Radio Network} ($CRN$) provides the capability of sharing the spectrum in an opportunistic manner by both licensed and unlicensed users, leading to an increase in the effective utilization of the available spectrum. According to a survey conducted by {\em Federal Communications Commission} ($FCC$)~\cite{b116, b117, b118, b119}, the usage of the radio spectrum is non-uniform. While some portions of the spectrum are heavily used, other portions remain relatively under-utilized. Thus, when a licensed user is not currently using the spectrum, an unlicensed user can sense this fact and may temporarily use this channel for his/her purpose. However, as soon as the licensed owner starts using his channel, the unlicensed user must relinquish this channel immediately, and move to a different one by sensing the {\em spectrum holes} or {\em white spaces}.

A cognitive radio should have the capability of being programmed to transmit and receive on a variety of frequencies and to use different transmission access technologies supported by its hardware design~\cite{b16,b23}. The transmission parameters, e.g., power level, modulation scheme, etc. of a cognitive radio should be reconfigurable not only at the beginning of a transmission but also during the transmission, when it is switched to a different spectrum band.

Of late, there also has been an increasing trend of multimedia communication in the form of voice, text, still image and video in various applications involving $CRN$. Designing efficient algorithms for allocating channels to a large number of such users of $CRN$s and maintaining the {\em Quality of Service} ($QoS$) for multimedia communication constitute an important research problem.

\subsection{Related Works} \label{related}

Multimedia communication through $CRN$s has already been studied by several authors~\cite{b31, b32, b29}. Mitola J. first introduced the concept of {\em flexible mobile multimedia communications}~\cite{b31} in a $CRN$. Kushwaha et al.~\cite{b32} used fountain coding for packet generation and conversion to send data with high reliability and tolerable delay. Shing et al.~\cite{b29} proposed the idea of dynamic channel selection for video streaming over a $CRN$, based on some priority-based scheduling of video signals. On the other hand Lei et. al. worked on spectrum fragmentation by their method ``Jello''~\cite{b111}, where they detect ``edges'' of power spectrum, then use classical best-fit, worst-fit and first-fit algorithm for spectrum selection and finally they do a distributed coordinate procedure to synchronize transceiver system. However, in all of these communication schemes, a video signal can not be communicated over the $CRN$ unless a channel of sufficiently large bandwidth for maintaining the $QoS$ of these video signals, is allocated from the white spaces of the spectrum. Thus, even if the sum of all the available white spaces in the spectrum may be larger than the required bandwidth for transmitting a video signal, it may not be possible to transmit the video signal if there is no single white space in the spectrum which can provide the required large bandwidth for its communication. Basically, this is a situation of {\em fragmentation} of the spectrum into small holes, with no hole being large enough to accommodate a video signal transmission. It was mentioned by Akyildiz et al.~\cite{b14} that {\em ``CR users may not detect any single spectrum band to meet the user's requirements. Therefore, multiple noncontiguous spectrum bands can be simultaneously used for transmission in CR networks''}. Some authors have addressed this implementation issue of the proposal by using {\em Orthogonal Frequency Division Multiplexing} ($OFDM$) - based $CRN$~\cite{b36, b37}. However, {\em Multi-Band $OFDM$} ($MBOFDM$) system for allowing more than one sender to send their messages in the $CRN$ is still a challenging problem~\cite{b35}.

Techniques for detection of unused spectrum and sharing the spectrum without harmful interference with other users with the help of a {\em Common Control Channel} ($CCC$) have been presented by Krishnamurthy et al.~\cite{b27}, Masri et al.~\cite{b28} and Bayhan and Alag\"{o}z~\cite{b109}. The $CCC$ is used for supporting the transmission coordination and spectrum related information exchange between the $CR$ users. It facilitates neighbor discovery, helps in spectrum sensing coordination, control signaling and exchange of local measurements between the $CR$ users. Spectrum sensing without using a $CCC$ has been considered by Kondareddy et al.~\cite{b5} and Xin and Cao~\cite{b6}.

Taxonomy, open issues, and challenges for channel assignment algorithms in a $CRN$ have been described in~\cite{b100}. Allocation schemes can be fixed~\cite{b30, b113}, dynamic~\cite{b1, b13, b29, b30, b113} or~\cite{b30, b113} depending on the flexibility of assigning channels to the cells in the network. The dynamic channel allocation in the spectrum is similar to the computer classical memory management strategies like ``first-fit'', ``best-fit'', and ``worst-fit''~\cite{b108}. Very recently, Bayhan and Alag\"{o}z~\cite{b109, b110} have proposed best-fit channel selection techniques in cognitive radio networks.

\subsection{Problem Statement} \label{problem}

\begin{figure}[]
  \centering
  \includegraphics[width=3.0in,height=0.25in]{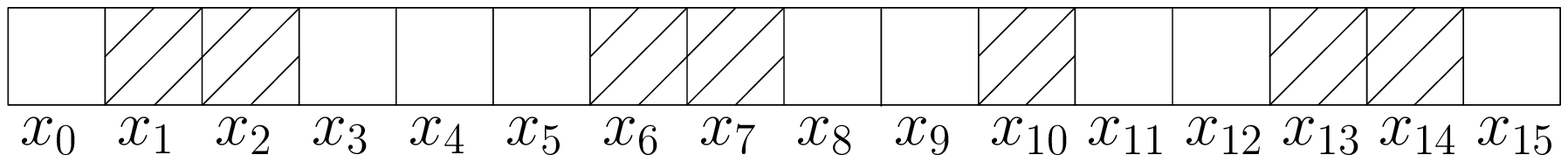}
  \caption{Spectrum Divided into Channels (Unused Channels shown as White)}\label{f3}
\end{figure}

Consider a representative scenario depicted in Fig.~\ref{f3} where we show a part of the spectrum divided into $16$ channels marked as $x_0, x_1, \cdots, x_{15}$, each of these channels being of the same bandwidth equal to $B_{min}$ which is the minimum bandwidth for a multimedia signal. For example, if the bandwidth requirements for the voice, text and video signals are $64$ Kbps, $128$ Kbps and $512$ Kbps respectively, then $B_{min}$ is taken to be $64$ Kbps. Thus, to transmit an audio signal, we need only one channel, while for a video signal, we need eight consecutive channels ($x_i$'s). However, as we see from Fig.~\ref{f3}, there is no continuous band consisting of eight channels, but a total of nine channels are still available. We need to devise an appropriate technique to allow the transmission of the given video signal through eight of these available nine channels, without compromising the video quality at the receiving end.

\begin{figure}[]
\centering
  \includegraphics[width=1.25in,height=.75in]{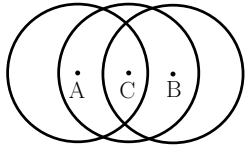}
  \caption{Nodes with their Respective Sensing Regions}\label{f6}
\end{figure}

Channel assignment process in a $CRN$ may broadly be divided in two subproblems - {\em channel sensing} and {\em channel allocation}. We assume that the transmission range of a node is equal to its sensing range. A node $U$ is called a 1-distance neighbor of a node $V$ if $U$ falls under the transmission or sensing range of the node $V$. While sensing, we assume that a node can always sense the channels which are being used by all of its 1-distance neighbors for transmitting their respective data. Referring to Fig.~\ref{f6}, the transmitting channels of all the neighbors at 1-distance from a node $A$ can be sensed by node $A$. Consider the node $C$ in Fig.~\ref{f6} which is a 1-distance neighbor of $A$. Node $B$ is another 1-distance neighbor of $C$ but node $B$ is at 2-distance apart from $A$. The channels used by $C$ in receiving some information from $B$, can not be sensed by node $A$. Thus, node $B$ can give rise to {\em hidden node problem}~\cite{b2} while allocating channels to node $A$. To be more specific, if $A$ and $B$ both want to communicate their messages to $C$ at the same time using the same channel (when both of them independently sense that channel as free), the node $C$ will experience a collision, and thus both the messages will be lost at $C$. The channel allocation algorithm must address this hidden node problem while allocating channels for the message communication from $A$ to $C$.

Another problem arises when, node $C$ has a capability of receiving a multimedia signal of bandwidth $512$ Kbps as shown in Fig.~\ref{f6}. Node $A$ sends some data to node $C$ which requires only $128$ Kbps bandwidth. Now, node $B$ also wants to send some data to node $C$ at the same time which requires, say, $384$ Kbps bandwidth. With the existing $OFDM$ technique~\cite{b34} in $CRN$, we can not transmit data simultaneously to the node $C$ from node $A$ and node $B$, though node $C$ might have the capacity to handle the data, unless there is a sufficient gap between the channels used for two different pairs of communicating nodes to avoid channel interference. According to Mahmoud et al.~\cite{b35}, the $MBOFDM$ system to handle such situation is a challenging problem due to synchronization requirement between the transmitter and the receiver.

We consider the situation for multimedia communication in which a typical user may require varying number of channels. Thus, a particular node may sometimes need just one single channel and sometimes a number of channels to communicate its messages depending on the types of the multimedia signals and their required $QoS$.

\subsection{Our Contribution} \label{contribution}

In this paper, we propose an elegant way of overcoming the problem of fragmentation of the available spectrum as mentioned in Section~\ref{problem}, with regard to the communication of multimedia signals over the $CRN$, while maintaining the $QoS$ requirement. We propose a technique for establishing a communication between sender and receiver nodes for single hop communication of multimedia data, where we first decompose a multimedia signal in time domain in terms of a number of bit-sets, with each set containing sufficiently sparsed bits so as to be transmitted over just a single channel of bandwidth $B_{min}$ and yet maintaining the signal quality. Thus, the total information content of a signal during a particular time frame is basically divided into several packets, with each packet being transmitted through one available channel in the white space. The constituent packets generated for a given time frame may, however, be transmitted over non-contiguous channels. At the receiver end, all the packets received through these channels will be used for reconstructing the original signal without degrading the signal quality.

Using the above strategy, we propose here a novel channel allocation technique which assigns non-contiguous channels for effecting multimedia communication between a sender and a receiver very fast through a random number generation process, so that the channel fragmentation problem as experienced in conventional first-fit or best-fit techniques is completely overcome resulting in very high channel utilization with negligible overhead in time. Detailed description of the proposed scheme is given below.

\begin{figure}[]
  \centering
  \includegraphics[width=3.0in,height=1.25in]{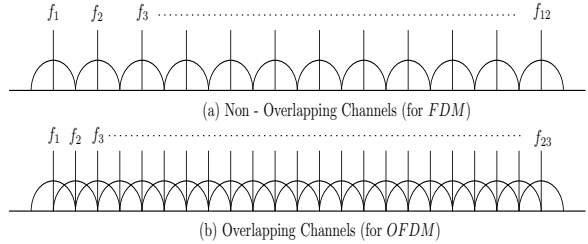}
  \caption{Channels configuration}\label{ONOC}
\end{figure}

To allow multiple senders for sending their data simultaneously through the $CRN$, we propose two possible channel allocation techniques, one based on {\em Frequency Division Multiplexing} ($FDM$) and {\em Frequency Division Multiple Access} ($FDMA$) ($FDM-FDMA$) and another based on  $OFDM$ and $FDMA$ ($OFDM-FDMA$). For the $FDM-FDMA$ allocation, we use the non-overlapping channels, where the channel width is assumed to be large enough to include the guard band, as shown in Fig.~\ref{ONOC}(a). Here, the basic idea is to use $FDM$ for every pair of communicating nodes, but $FDMA$ for different pairs of communicating nodes. Referring to Fig.~\ref{f6}, while the channels for communication between two nodes $A$ and $C$ are allocated using $FDM$ and the channels for communication between nodes $B$ and $C$ are also allocated using $FDM$, the channel allocation for the pairs $(A,C)$ and $(B,C)$ follows the $FDMA$ technique. For the  $OFDM-FDMA$ allocation, we use the overlapping orthogonal channels, as shown in Fig.~\ref{ONOC}(b). Here, the basic idea is to use $OFDM$ for every pair of communicating nodes, but $FDMA$ for different pairs of communicating nodes. To avoid inter-channel interferences we have to maintain certain minimum gap between every pair of channels allocated to different nodes. Referring to Fig.~\ref{f6}, while the channels for communication between two nodes $A$ and $C$ are allocated using $OFDM$ and the channels for communication between nodes $B$ and $C$ are also allocated using $OFDM$, the channel allocation for the pairs $(A,C)$ and $(B,C)$ follows the $FDMA$ technique. Thus, we have to maintain certain minimum gap between every pair of channels allocated to nodes $A$ and $B$ to avoid inter-channel interferences. We present algorithms for channel sensing, channel reservation and channel deallocation avoiding the hidden node problem and also avoiding possible collision with the channel demands from other users of the $CRN$. Corresponding transmission and reception protocols are also proposed.

We theoretically analyze our proposed algorithms to predict the average number of iterations or attempts made by our proposed algorithm for allocating the channels. In our later discussions, we use the terms {\em iterations} and {\em attempts} interchangeably throughout the text. The average number of such attempts is $O(1/f)$, where $f$ is the fraction of the free or available channels. In dynamic channel allocation, first-fit and best-fit techniques are commonly used ones~\cite{b109, b110, b111, b112, b39, b40}, and thus in our simulation, we compare our proposed protocol with first-fit and best-fit techniques for channel allocation. Simulation results show that the average number of attempts for acquiring the required number of channels agrees well to the theoretical values even for extremely heavy traffic with about $96\%$ blocked channels. From simulation, we also see that the proposed technique always outperforms the existing first-fit and best-fit~\cite{b109, b110} allocation techniques in terms of the average number of attempts needed for acquiring the necessary number of channels for all traffic situations ranging from light to extremely heavy traffic. The proposed technique can allocate the required numbers of channels in less than a second time with $FDM-FDMA$ even for $96\%$ traffic load and in less than $4.5$ sec with $OFDM-FDMA$ for $99\%$ traffic load, while the first-fit and best-fit techniques fail to allocate any channel in such situations. We can intuitively explain why our proposed technique performs better than the first-fit and best-fit techniques. Actually, both the latter techniques suffer from the channel fragmentation problem and channels cannot be allocated unless a contiguous free band of required number ($DN$) of channels is found. In contrast to this, our proposed technique removes the requirement of a single contiguous free band containing all these $DN$ channels and thus, the success rate is $100\%$ with our technique even with an extremely heavy traffic load when the existing approaches fail to allocate the channels. Moreover, because we are exploring the free channels through a random number generation process and every time we get a free channel, we include that one for our purpose, leads to a quick termination of our allocation algorithm with a success.

\section{Basic Ideas of our Proposed Protocol} \label{prelims}

\subsection{Creating Small Bandwidth Cognitive Sub-Data Channels}\label{CSBCUDC}

\begin{figure}[]
\centering
  \includegraphics[width=3.0in,height=1.5in]{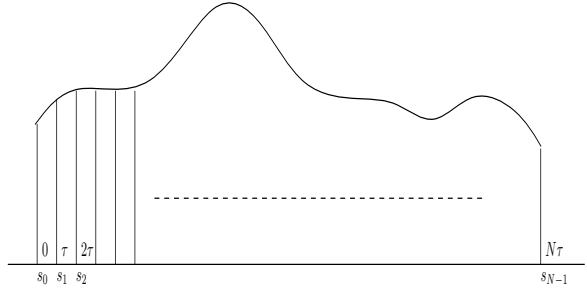}
  \caption{Samples from the signal}\label{f4}
\end{figure}

Consider a band-limited signal having a bandwidth of, say $W$. Let us assume that the signal is sampled with a sampling frequency of $2W$. Referring to Fig.~\ref{f4}, let $s_0, s_1, \cdots , s_{N-1}$ be $N$ samples taken over the time period $T$ of the band-limited signal at a sampling interval of $\tau = \frac{1}{2W}$. Thus, $T = N \tau$. Let us assume that from every sample, we get $b$ bits. Thus, the total number of bits is $N b$. So, the bits generated from all $N$ samples are $b_0, b_1, \cdots , b_{Nb-1}$. Let the bandwidth $W$ of this signal be less than or equal to $n B_{min}$. We then partition the $N b$ bits in $n$ subsets $BS_0, BS_1, BS_2, \cdots, BS_{n-1}$, where the bit-set $BS_i$ is defined as,
\begin{equation}
 BS_i = \{ b_j |j = i \mod n, 0 \le i, j \le n-1\}
\end{equation}
Note that in each of these $BS_i$'s, the bits are separated by $n \tau$ time, and hence, these would require a transmission bandwidth of $\frac {W} {n} \le B_{min}$. Thus, to transmit the original signal as shown in Fig.~\ref{f4}, we search for the availability of $n$ channels each of bandwidth $B_{min}$ in the white space of the spectrum.

\begin{figure}[]
\centering
  \includegraphics[width=3.0in,height=1.0in]{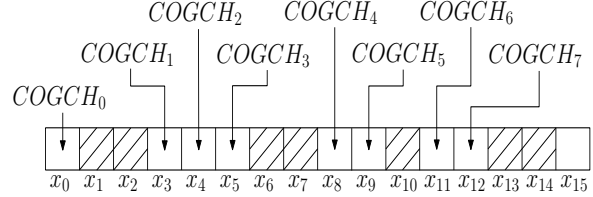}
  \caption{Utilized Spectrum with small bandwidth Cognitive Channels}\label{f5}
\end{figure}

Let $COGCH_i, i = 0, 1, \cdots, n-1$ be these $n$ cognitive channels such that the bits in the bit-set $BS_i$ is transmitted through $COGCH_i$ (as shown in Fig.~\ref{f5} for $n$ = 8). In practice, corresponding to each time frame of a suitable duration $T$, we take the bits in the bit-set $BS_i$ to form a data sub-packet $SP_i$. The header of each such sub-packet will contain the identity of the time frame (e.g., in the form of a packet number $PN$) as well as its {\em Sub-Packet Number} ($SPN$) equal to $i$. At the receiving end, all these received sub-packets having the same packet number will be used to reconstruct the original transmitted signal.

\subsection{Physical Implementation}\label{PIm}

\subsubsection{Physical Implementation of $FDM-FDMA$}\label{PIm1}

\begin{figure*}[]
\centering
\includegraphics[width=6.0in,height=2.50in]{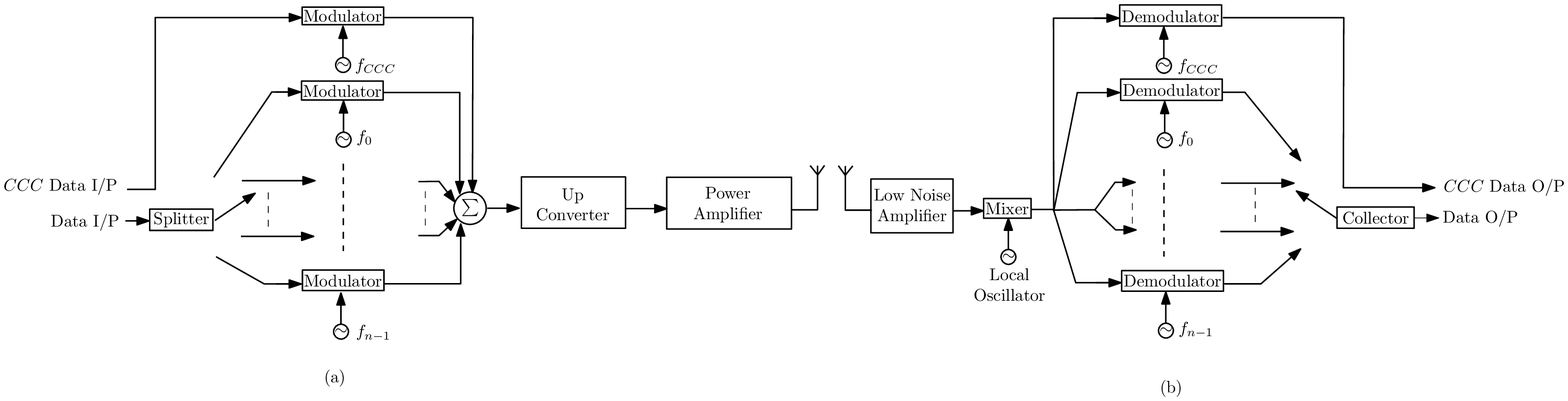}
\caption{(a) $FDM-FDMA$ Transmitter Block Diagram, (b) $FDM-FDMA$ Receiver Block Diagram}\label{Block}
\end{figure*}

We assume that all $CR$ users are {\em Secondary User}s ($SU$s) and have the same priority. Similarly, all {\em Primary User}s ($PU$s) are assumed to have the same priority which is greater than that of a $SU$. We also assume that any given node in the system has the maximum capability of providing some $DN$ ({\em Demand Number}) channels. Thus, a node $A$ may be allowed to be involved in simultaneously communicating more than one signal, so long as the sum of the numbers of channels used by it in communicating all these signals is less than or equal to $DN$. For example, voice ($64$ Kbps), data ($128$ Kbps), still image ($256$ Kbps), video ($384$ Kbps) and online streaming ($512$ Kbps) needs $DN$ as $1,~2,~4,~6$ and $8$, respectively, as we assume that $B_{min}$ is $64$ Kbps. We assume the presence of a dedicated $CCC$~\cite{b27,b28,b109} operating on a specific frequency ($f_{CCC}$) for coordination between the various $SU$s, with the communications through $CCC$ effected in discrete time slots. For a low traffic load, communication through $CCC$ can be done by following either $IEEE$ $802.11$ $CSMA/CA$~\cite{b114, b115} protocol. However, under moderate to heavy traffic, one may use any conventional {\em controlled access}~\cite{b114, b115} method like {\em Bit-Map}~\cite{b114, b115} protocol to improve the performance. If, we use {\em Bit-Map} protocol then each attempt made by our algorithm requires $O(\Delta)$ time, where $\Delta$ is the maximum node degree of the network.

The block diagrams of the proposed $FDM-FDMA$ transmitter and receiver have been shown in Figs.~\ref{Block}(a) and~\ref{Block}(b), respectively. In this scheme, for every channel we need a separate modulator and demodulator system. The $CCC$ channel, through which the control messages are transmitted, is totally separated from the data channels. The block {\em Splitter}, in Fig.~\ref{Block}(a), is working as a demultiplexer by which the $BS_i$ can be created, leading to generation of sub-packets $SP_i$. On the receiver side, the {\em Collector} in Fig.~\ref{Block}(b), is used to gather bits from different channels to form the packet constituted from the bits corresponding to all the $BS_i$s, which is required for regeneration of the multimedia data. Each $COGCH_i$ works on different frequencies $f_i$'s. For $FDM-FDMA$ technique, we can select all the frequencies $f_i$'s in non-overlapping channel (as shown in Fig.~\ref{ONOC}(a)). An alternative scheme using commercially available $MIMO$ system can also be used depending on the relative cost and ease of implementation.

\subsubsection{Physical Implementation of $OFDM-FDMA$}\label{PIm1}

\begin{figure}[]
  \centering
  \includegraphics[width=3.25in,height=1.25in]{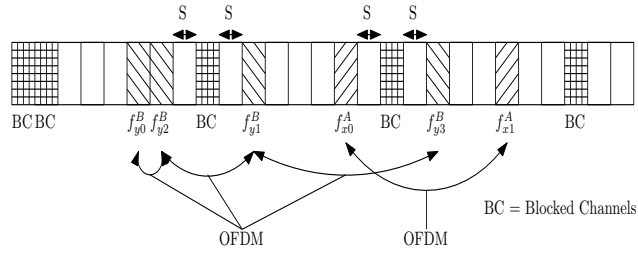}
  \caption{Channel allocation for node $A$ and node $B$ in overlapping channel} \label{chalo}
\end{figure}

For $OFDM-FDMA$ technique, the frequency $f_i$'s can be selected in such a way that all the $f_i$'s are orthogonal \cite{b34} leading to $OFDM$ modulation for one node which requires more than one channel to transmit its data, and the other nodes are to select free channels in such a way as to maintain certain minimum gap between two consecutively chosen channels to avoid inter-channel interference for different nodes in overlapping channels (as shown in Fig.~\ref{ONOC}(b)). As an example, referring to Fig.~\ref{f6}, nodes $A$ and $B$ need $2$ and $4$ channels respectively, to transmit some data to node $C$. Thus, node $A$ selects frequencies $f_{x0}^A$ and $f_{x1}^A$ for transmitting its data, while node $B$ selects frequencies $f_{y0}^B$, $f_{y1}^B$, $f_{y2}^B$ and $f_{y3}^B$ for its communication purpose as shown in Fig.~\ref{chalo}. Here, the carriers operating at $f_{x0}^A$ and $f_{x1}^A$ are orthogonal to each other and similarly $f_{y0}^B$, $f_{y1}^B$, $f_{y2}^B$ and $f_{y3}^B$ are also orthogonal to each other, but $f_{x0}^A$ and $f_{x1}^A$ need to be separated with some minimum band gap of $S$ (as shown in Fig.~\ref{chalo}) from the frequencies $f_{y0}^B$, $f_{y1}^B$, $f_{y2}^B$ and $f_{y3}^B$ to facilitate the synchronization process in the two destined receivers. Furthermore, to maintain the needed orthogonality condition between the $OFDM$ channels, all the $OFDM$ carriers need to be synchronously related to a unique pilot carrier, which can be transmitted (from some select nodes playing the role of collaborating leaders) in the same $OFDM$ band periodically over time. The frequency of the pilot carrier should be placed conveniently in the $OFDM$ frequency range (but not used by any node for data transmission). We, however, assume that this sacrifice of one single $OFDM$ carrier for the pilot in the entire $OFDM$ transmission bandwidth will not impact the spectral efficiency of the system at large. Every node will synthesize its own $OFDM$ carriers from the pilot carrier. In effect this will imply that, the $OFDM$ carrier frequencies transmitted from all the nodes and the pilot carrier frequency should be integrally related to a lower carrier frequency, which would be a highest common factor for all of them. This will lead to some additional hardware for the nodes along with the $IFFT/FFT$-based $OFDM$ generation and demodulation schemes with arbitrary but small number ($DN$) of $OFDM$ carriers. However, we consider this additional hardware complexity to be realizable with today's $VLSI$ design techniques and worthwhile as well keeping in mind the benefits that one would be able to derive in respect of the spectral efficiency achievable from this proposition.

\subsection{State Diagram of the Overall System}\label{SDOS}

\begin{figure}[]
\centering
\includegraphics[height=2.25in, width=3.0in]{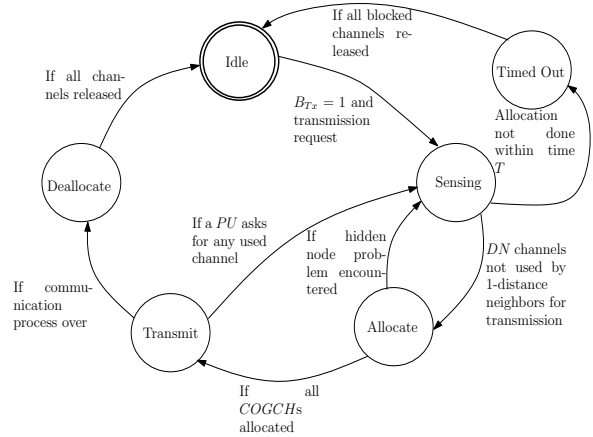}
\caption{State diagram}\label{f15}
\end{figure}

In Fig.~\ref{f15} we draw a state diagram that explains the basic functional units of the communication system as depicted through its various states and the state transition arcs. We start from the $Idle$ state. When the transmitter buffer of a node becomes full, a status bit $B_{Tx}$ of the node is set to $1$ (which is otherwise $0$). When $B_{Tx} = 1$ and this node wants to transmit, it moves from the $Idle$ state to the $Sensing$ state. The node in this $Sensing$ state explores the availability of free channels of the required number $DN$ as demanded by the multimedia signal to be transmitted by the node. If it finds $DN$ number of channels not being used by any of its 1-distance neighbors for transmission, then it blocks these channels temporarily and moves to $Allocate$ state. In $Allocate$ state, it determines whether there is any hidden node problem. If not, then it goes on to the $Transmit$ state, otherwise it moves back to the $Sensing$ state. In the $Sensing$ state, the node maintains a clock to measure the time needed for sensing and allocation of the required number $DN$ of channels. When the nodes first enters the $Sensing$ state from the $Idle$ state, the clock and the timing register both are set to $0$. The timing register is updated by the clock whenever the node moves from the $Allocate$ to $Sensing$ state. If allocation of channels is not done within a specified time-out period $T$, then the node moves to $Timed~Out$ state, at which it releases all the blocked channels, if any, and goes back to the $Idle$ state. In the $Transmit$ state, the node transmits its message through the allocated $DN$ channels. When the transmission is complete, it goes to $Deallocate$ state where it releases all these $DN$ number of blocked channels, resets $B_{Tx}$ to $0$ and goes back to the $Idle$ state. While the node is in the $Transmit$ state, if any primary user $PU$ reports, asking for any of the channels used by this node, then those channels will be immediately released and the system will go back to the $Sensing$ state, setting again both the clock and the timing register to $0$.

\section{Proposed Protocol}

\subsection{Algorithm for Connection Establishment} \label{connsetup}

Let $SA$ and $DA$ denote the addresses of the source node and the destination node, respectively. To establish a communication link the source node $SA$ needs to sense and allocate $DN$ number of channels for transmitting the multimedia signal using $CCC$. We consider below the connection establishment process for multimedia signals to be executed by the source node $SA$ and the destination node $DA$.

\begin{enumerate}
\item Sense the channels not being used by $A$'s 2-distance neighbors (to avoid the hidden node problem) as shown in Fig.~\ref{f6}. This would be effected with the help of some control and acknowledgement messages communicated through the $CCC$.

\item Allocate the $DN$ free channels found above to the destination node $C$ so that it becomes ready for receiving the desired multimedia signal from $A$.
\end{enumerate}

The above steps of allocating channels to any source-destination pair would be done {\em dynamically} in a {\em distributed manner} with the help of the $CCC$. After this allocation process, the actual multimedia communication between a source-destination pair will continue unless some or all of these channels are deallocated due to the arrival of one or more primary users.

\subsubsection{Reservation of Channel}\label{grabccc}

\begin{figure}[]  
  \centering
  \includegraphics[width=2in,height=0.25in]{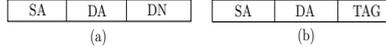}
  \caption{(a) $CM$ message and (b) $ACK$ or $WAIT$ message}\label{f13}
\end{figure}

The node $SA$ transmits a {\em Control Message} ($CM$) with $SA$, $DA$ and $DN$ values as shown in Fig.~\ref{f13}(a). After sending this control message, it waits up to some maximum time-out period, say $\delta_T$, for getting either an {\em Acknowledgement} ($ACK$) message or a $WAIT$ message from $DA$, both of which would contain the $SA$ and $DA$ values, with one more $TAG$ bit, as shown in Fig.~\ref{f13}(b), which is set to '$0$' for an $ACK$ message and '$1$' for a $WAIT$ message. The $ACK$ message is sent if node $DA$ is capable of providing $DN$ number of channels for receiving the multimedia signal from $SA$ (i.e., when the available number of channels $AN$ at $DA$ is greater than or equal to $DN$, while the $WAIT$ message is sent when $AN \le DN$.) Since the node $DA$ may simultaneously receive such channel reservation requests from other source nodes as well, for $AN \geq DN$, it temporarily reserves the requested number (i.e., $DN$) of channels for node $SA$ on a first-come-first-serve basis (without bothering about which $DN$ channels). If the node $DA$ is not capable of allocating the requested $DN$ number of channels to $SA$, then along with sending the $WAIT$ message to the node $SA$, it puts this request from $SA$ (in the form of $CM$) in a waiting queue for later servicing. If neither the $ACK$ nor the $WAIT$ message is received by $SA$ within $\delta_T$ time (due to a possible collision caused by the simultaneous transmission of messages from some other node(s) within 1-distance from $SA$ or due to the hidden node problem, i.e., due to a collision at the node $DA$ caused by messages from some node, say $V$ which is at 1-distance from $DA$, but at 2-distance from $SA$), then $SA$ retransmits its control message $CM$. This process of retransmission is repeated by $SA$ until an $ACK$ or $WAIT$ message is received from $DA$. The algorithms {\em reserve\_channels\_transmitter} and {\em reserve\_channels\_receiver} to be executed by nodes $SA$ and $DA$ are given in Algorithm~\ref{algo1} and~\ref{algo2} respectively.

\begin{algorithm}[]
\scriptsize
\linesnumbered
\KwIn{$SA$, $DA$ and $DN$}
\KwOut{channels\_reserved}
$channels\_reserved = false$ \;
\While{$channels\_reserved = false$ AND $B_{Tx} = 1$} {
  Transmit $CM$ to the node $DA$\;
  Wait for $\delta_T$ time to receive $ACK$ or $Wait$ Signal\;
  \If{$ACK$ or $WAIT$ Signal received within $\delta_T$}
  {
    \eIf{$ACK$ is received within $\delta_T$}
    {
      channels\_reserved = true\;
    }
    {
      Wait for $ACK$ from $DA$ /* $WAIT$ received */\;
      channels\_reserved = true\;
    }
  }
}
\caption{{\em Reserve\_Channels\_Transmitter}} \label{algo1}
\normalsize
\end{algorithm}

\begin{algorithm}[]
\scriptsize
\linesnumbered
\KwIn{$CM$ (consisting of $SA$, $DA$ and $DN$)}
\KwOut{$ACK$, $WAIT$ to the Transmitter}
\eIf{$AN$ $\geq$ $DN$}
{
  $AN=AN-DN$\;
  Update its database\;
  Transmit $ACK$\;
}
{
  Enter $CM$ in a waiting queue\;
  Send back $WAIT$ Signal to node $SA$\;
}
\caption{{\em Reserve\_Channels\_Receiver}} \label{algo2}
\normalsize
\end{algorithm}

\subsubsection{Sensing and Allocation of Channels} \label{SAC}

\paragraph{Sensing and Allocation of Channels for $FDM-FDMA$ Technique} \label{SAC11}

After getting the $ACK$ message from the destination node $DA$ in reply to the $CM$ message as described in Section~\ref{grabccc}, the source node $SA$ will try to find the required $DN$ number of data channels from the currently available white spaces of the spectrum. This will be done by randomly choosing a set of $DN$ distinct channels which are not being used by any of the 2-distance neighbors of $SA$ for transmission as well as reception (to avoid the hidden node problem) of data. We assume that the width of the non-overlapping channel already includes the bandgap to avoid inter-channel interference and hence, a free channel for a node $U$ means one which is not being used by any other node within the 2-distance neighborhood of $U$. We randomly generate a number $i$ and then sense whether channel $i$ is free (with respect to all nodes in 2-distance neighborhood of $U$). If channel $i$ is free, then it can be allocated to $U$. In fact, we generate $DN$ such random numbers and sense in parallel whether all the channels corresponding to these randomly generated $DN$ numbers are free. If $m$ free channels are found by this process, then the process terminates if $m = DN$; otherwise the whole process is repeated to find the required number $m-DN$ of free channels to be allocated to $U$. Each iteration of this loop is termed as an {\em attempt}, as introduced earlier in Section~\ref{contribution}.

\begin{figure}[]
  \centering
  \includegraphics[width=3.0in,height=0.5in]{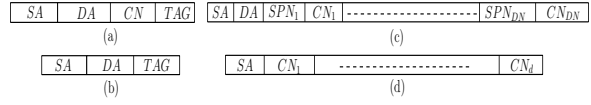}
  \caption{(a) $TAM$ or $CCB$ message, (b) $TAM\_ACK$ or $NACK$ message, (c) $CHALLOC$ message and (d) $CRM$ message ($d=|channel\_set|$)}\label{f14}
\end{figure}

\begin{figure}[]
  \centering
  \includegraphics[width=2.5in,height=1.25in]{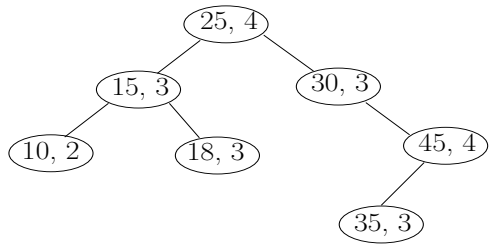}
  \caption{$AVL$ tree for storing the channel usage status of a node}\label{f17}
\end{figure}

The fact that none of the 1-distance neighbors of $SA$ is currently using a given channel for transmitting their data, can easily be determined by listening to this channel by $SA$ (channel sensing). However, to avoid the hidden node problem, whether a given channel is being used by any node $U$ among the 1-distance neighbors of $SA$ for receiving some messages from some node, say $V$, which is at 2-distance from $SA$, can not be determined by such channel sensing. To decipher that, node $SA$ has to send a {\em Trial Allocation Message} ($TAM$) to all of its 1-distance neighbors which would contain the source and destination addresses (i.e., $SA$ and $DA$) along with the {\em Channel Number} ($CN$) in question. The structure of the $TAM$ message is as shown in Fig.~\ref{f14}(a), where the $TAG$ field is set to '$0$' for a $TAM$. On getting this $TAM$, a node $U$ would send back an acknowledgement ($TAM\_ACK$) or a no-acknowledgement ($NACK$) message to $SA$ depending on whether $U$ is currently not using the channel $CN$ for receiving any message or not, respectively. The $TAM\_ACK$ and $NACK$ messages are of the form as shown in Fig.~\ref{f14}(b), where the $TAG$ field is set to '$00$' for a $NACK$ and '$01$' for an $TAM\_ACK$. Node $U$ can check this fact efficiently if it maintains a channel usage database in the form an $AVL$ tree as shown in Fig.~\ref{f17} where each node of the tree contains a tuple ($CN$, $SA$) and insertion or finding an element in the tree is done based on the $CN$ field only. The choice of $AVL$ tree as the data structure for this purpose enables us insertion, deletion and finding an element from it all in $O(log~m)$ time where $DN$ is the total number of nodes in this tree. In case $U$ is currently not using the channel $CN$ for receiving any message, it temporarily allocates the channel $CN$ to the node $SA$ and keeps this information by inserting a new node with this $CN$ and $SA$ information in the $AVL$ tree. This would help prohibiting other nodes selecting this channel $CN$ for transmitting their data when the node $SA$ is still in the process of selecting all of its required channels and has not yet completed that process. If $SA$ does not receive any $NACK$ message within a maximum time-out period $\delta_T$ from any node in reply to this $TAM$ message, then $SA$ puts this channel number $CN$ in its chosen set of channels $channel\_set$; otherwise, $SA$ can not use the channel $CN$ for transmitting its data and hence it broadcasts a {\em Clear Channel Blockage} ($CCB$) message to all of its 1-distance neighbors. The structure of the $CCB$ message is same as that of a $TAM$ shown in Fig.~\ref{f14}(a), where the $TAG$ field is set to '$1$' for a $CCB$. If a node $U$ receives this $CCB$ message, then $U$ will delete the corresponding node from its $AVL$ tree storing its channel usage status (thus, the channel $CN$ will now be treated as available by the node $U$).

The above process of allocating channels for $SA$ will be repeated to get all $DN$ channels after which the transmission will be started. When the required number of channels are found through the above process, a $Channel~Allocate~(CHALLOC)$ command is broadcast by $SA$ to its 1-distance neighbors with the information regarding the destination node $DA$, and the sub-packet number ($SPN$) of every packet along with the allocated channel number ($CN$) as shown in Fig.~\ref{f14}(c). On receiving this $CHALLOC$ command, node $DA$ will record the information regarding the $SPN$ and $CN$ for the sub-packets to be received from $SA$ in its channel reservation database, while any other node will release the temporary blockage of the corresponding channel numbers. If, however, the required number of channels are not found within a maximum time unit, say $T$ ($\delta_T \ll T$), then the node $SA$ can not start its transmission at the moment and it broadcasts a $Channel~ Release~ Message$ ($CRM$) signal of the form as shown in Fig.~\ref{f14}(d), to all of its 1-distance neighbors to release temporarily blocked channels. Node $SA$ has to try again for getting the required $DN$ number of channels until success or the transmitter buffer becomes $0$. The algorithms {\em Sense\_Allocate\_Transmitter\_$FDM-FDMA$} and {\em Sense\_Allocate\_Receiver} to be executed by the node $SA$ and any other receiving node are given in Algorithm~\ref{algo3}, and~\ref{algo6}, respectively.

\begin{algorithm}[]
\scriptsize
\linesnumbered
\KwIn{$DA$, $DN$, $MAX$}
\KwOut{Selected channel numbers $c_1, c_2, \cdots, c_{DN}$}
$channel\_set = \emptyset$\;
$cardinality = 0$     $/*cardinality = |channel\_set|*/$\;
$j = DN$\;
$time = 0$\;
\While{$time$ $\le$ $T$} {
  Randomly generate a set of $j$ distinct channel numbers $c_1, c_2, \cdots, c_j$ in the range $1$ to $MAX$ such that for $1 \leq i \leq j$, $c_i \notin channel\_set$ \;
  Sense channel numbers $c_1, c_2, \cdots, c_j$ in parallel        /*to check if channel $c_i$ is idle*/\;
  \For{$i = 1$ to $j$}
  {
    \If{$|channel\_set| < DN$}
    {
      \If{channel $c_i$ idle}
      {
	Form the $TAM$ message with $CN = c_i$\;
	$trial\_no\_TAM= 1$\;
	$flag = true$\;
	\While{($trial\_no\_TAM < Max\_trial\_TAM$ AND $flag$)} {
	  Broadcast the $TAM$ message to all 1-distance neighbors using the $CCC$\;
	  Wait up to a maximum time of $\delta_T$ to receive reply message(s)\;
	  \eIf{reply received}
	  {
	    flag=false\;
	    \eIf{no $NACK$ received}
	    {
	      $channel\_set = channel\_set \cup \{c_i\}$\;
	      $cardinality=cardinality+1$\;
	    }
	    {
	      Send $CCB$ with $CN$\;
	    }
	  }
	  {
	    $trial\_no\_TAM = trial\_no\_TAM + 1$\;
	  }
	}
      }
    }
  }
  $time = time + 1$\;
}
\eIf{$DN = cardinality$}
{
  Broadcast $CHALLOC$ command formed with the $channel\_set$ to all 1-distance neighbors\;
}
{
  Broadcast $CRM$ formed with the $channel\_set$ to all 1-distance neighbors to release all temporarily blocked channels\;
}
\caption{{\em Sense\_Allocate\_Transmitter\_$FDM-FDMA$}} \label{algo3}
\normalsize
\end{algorithm}

\begin{algorithm}[]
\scriptsize
\linesnumbered
\KwIn{$TAM$, $CHALLOC$}
\KwOut{Select and locked data channels.} /* The following code will be executed by all nodes receiving the $TAM$ and $CHALLOC$ messages*/ \;
\If{$TAM$ received with source node $SA$ and channel number $CN$} {
  \eIf{channel $CN$ is free for its 1-distance neighbor AND $CN$ is not temporarily blocked for any other node}
  {
    Update its channel usage database by temporarily marking channel number $CN$ as being used by node $SA$\;
    Transmit $TAM\_ACK$ to $SA$ through $CCC$\;
  }
  {
    Transmit $NACK$ to $SA$ through $CCC$\;
  }
}
\If{$CHALLOC$ received}
{
  \eIf {$DA$ in $CHALLOC$ = its own $id$}
  {
    Update its channel reservation database with $SPN$ and $CN$ values assigned from $CHALLOC$ for each channel\;
  }
  {
    Update its channel usage database by releasing the temporarily blocked channel numbers indicated in $CHALLOC$\;
  }
}
\If{$CRM$ received}
{
  Update its channel usage database by releasing the temporarily blocked channel numbers indicated in $CRM$\;
}
\caption{{\em Sense\_Allocate\_Receiver}} \label{algo6}
\normalsize
\end{algorithm}

\paragraph{Sensing and Allocation of Channels for $OFDM-FDMA$ Technique} \label{SAC1}

\begin{algorithm}[]
\scriptsize
\linesnumbered
\KwIn{$DA$, $DN$, $MAX$}
\KwOut{Selected channel numbers $c_1, c_2, \cdots, c_{DN}$}
$channel\_set = \emptyset$\;
$j = DN$\;
$time = 0$\;
$flag_1=true$\;
$required\_channel\_numbers=DN$\;
\While{$time$ $\le$ $T$ AND $flag_1$} {
  $temp\_set = \emptyset$\;
  Randomly generate a set of $j$ distinct channel numbers $c_1^1, c_2^1, \cdots, c_j^1$ in the range $1$ to $MAX$ such that for $1 \leq i \leq j$, $c_i^1 \notin channel\_set$ \;
  \For{$k=1$ to $j$}
  {
    $c_k^2 = c_k^1 + 1$ \;
    $c_k^3 = c_k^1 + 2$ \;
  }
  Create a list of $3 \times DN$ numbers with $c_1^1, c_1^2, c_1^3 \cdots, c_j^1, c_j^2, c_j^3$\;
  Sense channel numbers $c_1^1, c_1^2, c_1^3 \cdots, c_j^1, c_j^2, c_j^3$ in parallel        /*to check if channel $c_i^1, c_i^2, c_i^3$ is idle*/\;
  \For{$i = 1$ to $j$}
  {
    \For{$k=1$ to $3$}
    {
      \If{channel $c_i^k$ idle}
      {
	Form the $TAM$ message with $CN = c_i^k$\;
	$trial\_no\_TAM= 1$\;
	$flag_2 = true$\;
	\While{($trial\_no\_TAM < Max\_trial\_TAM$ AND $flag_2$)} {
	  Broadcast the $TAM$ message to all 1-distance neighbors using the $CCC$\;
	  Wait up to a maximum time of $\delta_T$ to receive reply message(s)\;
	  \eIf{reply received}
	  {
	    $flag_2=false$\;
	    \eIf{no $NACK$ received}
	    {
	      $temp\_set = temp\_set \cup \{c_i^k\}$\;
	    }
	    {
	      Send $CCB$ with $CN$\;
	    }
	  }
	  {
	    $trial\_no\_TAM = trial\_no\_TAM + 1$\;
	  }
	}
      }
    }
  }

  $find\_free\_bands(temp\_set,required\_channel\_numbers,$ $channel\_set)$\;

  \If{$required\_channel\_numbers=0$}
  {
    $flag_1=false$\;
  }
  $time = time + 1$\;
}
\eIf{$DN = |channel\_set|$}
{
  Broadcast $CHALLOC$ command formed with the $channel\_set$ to all 1-distance neighbors\;
}
{
  Broadcast $CRM$ formed with the $channel\_set$ to all 1-distance neighbors to release all temporarily blocked channels\;
}
\caption{{\em Sense\_Allocate\_Transmitter\_$OFDM-FDMA$}} \label{algo4}
\normalsize
\end{algorithm}

\begin{algorithm}[]
\scriptsize
\linesnumbered
\KwIn{$temp\_set$, $required\_channel\_numbers$\;}
\KwOut{$required\_channel\_numbers$, $channel\_set$}

  Sort $temp\_set$ in non-increasing order\;
  Scan $temp\_set$ to form $band\_set$ with the 2-tuple ($band\_length$, $start\_channel\_number$) as its element \;
  Set $t \leftarrow |band\_set|$  \;
  Sort $band\_set$ in non-increasing order of $band\_length$ component of its elements (2-tuples) \;

  \For{$i=1$ to $t$}
  {
    \If{$band\_set[i].band\_length>3$}
    {
      $channel\_set = channel\_set \cup \{band\_set[i].start\_channel\_number+1, \cdots, band\_set[i].start\_channel\_number+band\_set[i].band\_length-2 \}$\;
      $temp\_number= min[DN, band\_set[i].start\_channel\_number+band\_set[i].band\_length-2]$ \;
      $required\_channel\_numbers=required\_channel\_numbers-temp\_number$\;
    }
  }
  return($required\_channel\_numbers$, $channel\_set$);
\caption{{\em procedure $find\_free\_bands$}} \label{algo5}
\normalsize
\end{algorithm}

Node $SA$ has to send a {\em Trial Allocation Message} ($TAM$) to all of its 1-distance neighbors which would contain the source and destination addresses (i.e., $SA$ and $DA$) along with the {\em Channel Number} ($CN$) in question. The structure of the $TAM$ message is as shown in Fig.~\ref{f14}(a), where the $TAG$ field is set to '$0$' for a $TAM$. On getting this $TAM$, a node $U$ would send back an acknowledgement ($TAM\_ACK$) or a no-acknowledgement ($NACK$) message to $SA$ depending on whether $U$ is currently not using the channel $CN$ for receiving any message or not, respectively. The $TAM\_ACK$ and $NACK$ messages are of the form as shown in Fig.~\ref{f14}(b), where the $TAG$ field is set to '$00$' for a $NACK$ and '$01$' for an $TAM\_ACK$. Node $U$ can check this fact efficiently if it maintains a channel usage database in the form an $AVL$ tree as shown in Fig.~\ref{f17} where each node of the tree contains a tuple ($CN$, $SA$) and insertion or finding an element in the tree is done based on the $CN$ field only. The choice of $AVL$ tree as the data structure for this purpose enables us insertion, deletion and finding an element from it all in $O(log~m)$ time where $DN$ is the total number of nodes in this tree. In case $U$ is currently not using the channel $CN$ for receiving any message, it temporarily allocates the channel $CN$ to the node $SA$ and keeps this information by inserting a new node with this $CN$ and $SA$ information in the $AVL$ tree. This would help prohibiting other nodes selecting this channel $CN$ for transmitting their data when the node $SA$ is still in the process of selecting all of its required channels and has not yet completed that process. If $SA$ does not receive any $NACK$ message within a maximum time-out period $\delta_T$ from any node in reply to this $TAM$ message, then $SA$ puts this channel number $CN$ in its chosen set of channels $channel\_set$; otherwise, $SA$ can not use the channel $CN$ for transmitting its data and hence it broadcasts a {\em Clear Channel Blockage} ($CCB$) message to all of its 1-distance neighbors. The structure of the $CCB$ message is same as that of a $TAM$ shown in Fig.~\ref{f14}(a), where the $TAG$ field is set to '$1$' for a $CCB$. If a node $U$ receives this $CCB$ message, then $U$ will delete the corresponding node from its $AVL$ tree storing its channel usage status (thus, the channel $CN$ may now be treated as available by the node $U$).

We assume that a free overlapping channel to be used by a node $U$ refers to a channel which is i) not being used by any node within the 2-distance neighborhood of $U$ and ii) sufficiently separated from those channels which are being used by all nodes in the 2-distance neighborhood to avoid inter-channel interference. We assume a gap of one channel on either side of the channel to be used by $U$ to avoid this interference. Thus, if channel $i$ is to be allocated to $U$, then channels $i-1$, $i$ and $i+1$ must not be used by any node in the 2-distance neighborhood of $U$. However, it can be generalized and we assume that with $OFDM$ communication, a contiguous set of $m$ channels, $1 \leq m \leq DN$ may be allocated to a node $U$, provided we find a set of $m+2$ contiguous channels which are not being used by any of the nodes in the 2-distance neighborhood of $U$. Thus, if none of the channels $i-1, ~i, ~\cdots~, ~i + m$ are being used by any node in the 2-distance neighborhood of $U$, then the $m$ channels $i, ~i+1, ~\cdots~, ~i+m-1$ can be used by $U$ in the $OFDM$ mode. After randomly generating $i$, we sense whether channels $i$, $i+1$ and $i+2$ are free (with respect to all nodes in 2-distance neighborhood of $U$). We mark all these free channels. Here also, we generate $DN$ such random numbers and sense in parallel whether all the channels corresponding to these randomly generated $DN$ numbers are free, and mark all these free channels found by this step. After this, we check the status of all channels to find a consecutive band of $m$ free channels, $3 < m < DN+2$, out of which $(m-2)$ consecutive channels may be allocated to $U$. If $m - 2 = DN$, then the process is terminated; otherwise, the whole process is repeated for finding the $m - 2- DN$ channels still to be allocated to $U$. As with $FDM-FDMA$ implementation, one iteration of this loop is termed as an {\em attempt}. Thus, the sensing time per attempts in $OFDM-FDMA$ channel allocation technique is three times more than that in $FDM-FDMA$ channel allocation technique.

The detailed steps for finding free bands after getting the free channel numbers have been presented in the {\em procedure $find\_free\_bands$} (Algorithm~\ref{algo5}). First, the free channels numbers are included in a set $temp\_set$ which is then sorted in non-increasing order. This sorted $temp\_set$ is then scanned once from left to right to produce a set $band\_set$ containing the 2-tuples ($band\_length$, $start\_channel\_number$) as its elements. This set $band\_set$ is then sorted in non-increasing order based on the $band\_length$ field of each 2-tuple. Finally, this sorted $band\_set$ is scanned once from largest to smallest $band\_length$ to collect the free bands with largest possible sizes to form the $channel\_set$. Since the number of elements in $temp\_set$ is small (less than $3 \times DN$), the total time for executing this procedure will be very small.

The above process of allocating channels for $SA$ will be repeated to get all $DN$ channels after which the transmission will be started. When the required number of channels are found through the above process, a $Channel~Allocate~(CHALLOC)$ command is broadcast by $SA$ to its 1-distance neighbors with the information regarding the destination node $DA$, and the sub-packet number ($SPN$) of every packet along with the allocated channel number ($CN$) as shown in Fig.~\ref{f14}(c). On receiving this $CHALLOC$ command, node $DA$ will record the information regarding the $SPN$ and $CN$ for the sub-packets to be received from $SA$ in its channel reservation database, while any other node will release the temporary blockage of the corresponding channel numbers. If, however, the required number of channels are not found within a maximum time unit, say $T$ ($\delta_T \ll T$), then the node $SA$ can not start its transmission at the moment and it broadcasts a $Channel~ Release~ Message$ ($CRM$) signal of the form as shown in Fig.~\ref{f14}(d), to all of its 1-distance neighbors to release temporarily blocked channels. Node $SA$ has to try again for getting the required $DN$ number of channels until success or the transmitter buffer becomes $0$.

The algorithm {\em Sense\_Allocate\_Transmitter\_$OFDM-FDMA$} to be executed by the node $SA$ and any other receiving node are given in \ref{algo4}.

\subsection{Algorithms for Transmission and Reception}\label{datacomm}

\begin{figure}[]  
  \centering
  \includegraphics[width=1.25in,height=0.15in]{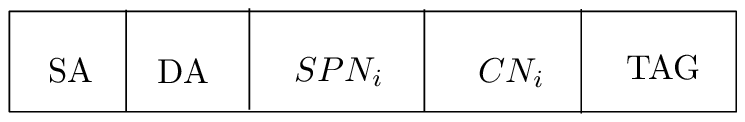}
  \caption{$DATA\_ACK$ message}\label{f20}
\end{figure}

When all the required $DN$ channels are allocated to both the nodes $SA$ and $DA$, the node $SA$ starts transmission of its multimedia data following the algorithm {\em Transmit\_Data\_Packet} given below. The receiving node $DA$ will execute the algorithm {\em Receive\_Data\_Packet} described below to receive the $DN$ sub-packets corresponding to each sub-packet number $SPN$ and will reconstruct the original message from these sub-packets. If a sub-packet is received correctly by $DA$, then an acknowledgement message ($DATA\_ACK$) will be sent by $DA$ back to $SA$. The structure of the $ACK$ message is as shown in Fig.~\ref{f20}. If $DATA\_ACK$ is not received within time out period $\delta_T$, then node $SA$ has to sense if a primary user has started using his channel. Then it immediately relinquishes this channel. $SA$ will then look for some other alternative channel which can be allocated for transmitting the corresponding data sub-packet. If this is not possible in an extreme situation with a maximum number of trials, say $maxtrial$, then the node $SA$ has to abort the transmission. The algorithms {\em Transmit\_Data\_Packet} and {\em Receive\_Data\_Packet} to be executed by nodes $SA$ and $DA$ are given in Algorithms~\ref{algo7} and~\ref{algo8} respectively.

\begin{algorithm}[]
\scriptsize
\linesnumbered
\KwIn{$DN$, $channel\_set$, packet to be transmitted, $maxtrial$}
\KwOut{Transmitted packets}
$abort = false$\;
$PN = 0$\;
\While{$B_{Tx} = 1$ AND $abort = false$} {
  \For{$i=0$ to $DN - 1$}
  {
    Form the sub-packet $SPN_i$ with packet number = $PN$,
    $sub\_packet\_number = i$\;
    $sub\_packet\_received[i] = false$\;
  }
  \ForAll{$COGCH_i$, $0 \leq i \leq (DN - 1)$, in parallel}
  {
    $trial\_number = 1$\;
    \While{$trial\_number \le maxtrial$ AND $sub\_packet\_received[i] = false$}
    {
      Transmit the sub-packet $SPN_i$ through the channel $COGCH_i$\;
      \eIf{$DATA\_ACK$ received within the time out period $\delta_T$}
      {
        $sub\_packet\_received = true$\;
      }
      {
	Sense if $PU$ uses this channel\;
	\If {$PU$ uses this channel}
	{
	  Release this channel and look for another available channel using Algorithm~\ref{algo3} ($FDM-FDMA$) or~\ref{algo4} ($OFDM-FDMA$) \;
	  \If{a new channel number $new\_channel$ is found}
	  {
	    $COGCH_i = new\_channel$\;
	    $trial\_number = 1$ \;
	    /*re-transmission of $sub\_packet[i]$ is started on this $new\_channel$ */\\
	  }
	}
      }
      $trial\_number = trial\_number + 1$\;
    }
    \If {$sub\_packet\_received = false$}
    {
      $abort = true$\;
    }
  }
  $PN = PN + 1$\;
}
\caption{{\em Transmit\_Data\_Packet}} \label{algo7}
\normalsize
\end{algorithm}

\begin{algorithm}[]
\scriptsize
\linesnumbered
\KwIn{Received Packet from Transmitter}
\KwOut{$DATA\_ACK$ messages to the transmitting node $SA$} /* to be executed by the receiving node $DA$ */ \;
\ForAll{$COGCH_i$, $0 \leq i \leq (DN - 1)$, in parallel} {
  \If{packet received correctly with packet number $PN$}
  {
    Send $DATA\_ACK$ message to the transmitting node $SA$ with packet number $PN$ and sub-packet number $i$\;
  }
}
\caption{{\em Receive\_Data\_Packet}} \label{algo8}
\normalsize
\end{algorithm}

\subsection{Algorithm for Deallocation of Channels}\label{Deallo}

\begin{figure}[]  
  \centering
  \includegraphics[width=1.5in,height=0.15in]{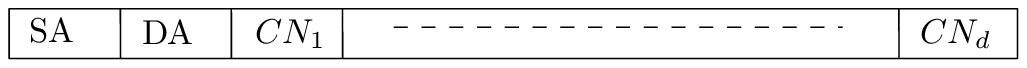}
  \caption{$CLS$ message}\label{f21}
\end{figure}

After successful transmission of all of its data packets, the transmitting node $SA$ will release all the data channels used by it (by deleting the corresponding entries from its $AVL$ tree storing the channel usage status). Also it issues a channel release message clear signal ($CLS$) of the form shown in Fig.~\ref{f21} through $CCC$. The receiving node $DA$ release all data channels used by node $DA$ for this communication (update $AVL$ tree) and update its $AN$. All other 1-distance neighbors are also deleting the corresponding entries from its $AVL$ tree storing the channel usage status. In case the node $SA$ has to abort a transmission, it releases all the channels allocated to both $SA$ and $DA$ in the same way. When one or more channels used by the node $DA$ are released, the next channel reservation request from its waiting queue is considered if that can be satisfied. The waiting queue can be implemented using a linked list with $INFO$ field of each node containing the $2$-tuple ($SA$, $DN$). However, sensing these waiting requests in a {\em First-Come-First-Serve} ($FCFS$) order may result in a poor utilization of the channels. Instead, some other variants of this servicing policy may be chosen to increase the channel utilization. For example, the request from a node with the minimum number of required channels from amongst those waiting for the service may be chosen. This would increase the channel utilization, but in turn, may lead to starvation (similar to {\em Shortest-Job-First} ($SJF$) CPU scheduling in operating systems~\cite{b33}) of the requests with a large value of $DN$. This problem of starvation may, however, be avoided by taking into account the ageing factor of the accumulated requests, resulting into an increased channel utilization with no starvation. The algorithms {\em Deallocate\_Data Channels\_Transmitter} and {\em Deallocate\_Data Channels\_Receiver} to be executed by the node $SA$ and $DA$ are given in Algorithms~\ref{algo9} and~\ref{algo10} respectively.

\begin{algorithm}[]
\scriptsize
\linesnumbered
\KwIn{Transmission completion signal}
\KwOut{Deallocation of all channels}
\If{ data transmission completed}
{
  Set $B_{Tx}=0$\;
}

\For{$COGCH_i \mid_{(0\le i\le n-1)}$}
{
  Transmits $CLS$ to all 1-distance neighbors through $CCC$\;
  Release all data channels\;
}
\caption{Deallocate\_Data Channels\_Transmitter.} \label{algo9}
\normalsize
\end{algorithm}

\begin{algorithm}[]
\scriptsize
\linesnumbered
\KwIn{$CLS$ from transmitter}
\KwOut{Deallocation of all channels}
\For{$COGCH_i \mid_{(0\le i\le n-1)}$}
{

  \If{$CLS$ received}
  {
    Release all data channels\;
  }
  \If{$DA$ in $CLS$ = its own id}
  {
    Update $AN$\;
    Process the waiting queue\;
  }
}
\caption{Deallocate\_Data Channels\_Receiver.} \label{algo10}
\normalsize
\end{algorithm}

\section{Performance Analysis}\label{PA}

\subsection{Performance of Channel Allocation Algorithm using $FDM-FDMA$ Technique}\label{PCAA}

Let $C$ be the total number of channels out of which we assume that $\pi$ channels are in the primary band and the rest are in the secondary band. At any time instant $t$, let $B_{p,t}$ and $B_{s,t}$ be the numbers of blocked (already allocated by 2-distance neighbors and maintain certain minimum gap between two consecutively chosen channels to avoid inter-channel interference for different nodes) channels in the primary band and the secondary band, respectively. Thus, the total number of blocked channels at time $t$ is given by $B_t = B_{p,t} + B_{s,t}$. Let $F_{p,t}$ be the number of free channels in the primary band at time $t$, which is given by $\pi-B_{p,t}$. Similarly, let $F_{s,t}$ be the number of free channels in the secondary band at time $t$, which is given by $C-\pi-B_{s,t}$. Let $F_t = F_{p,t} + F_{s,t}$. Let there be a request at time $t$ for allocating $n$ channels to communicate a given multimedia signal. Referring to Algorithm~\ref{algo3}, we try to reserve the required number of channels, i.e., $n$ channels in successive attempts, where each attempt corresponds to a single execution of steps $5$ to $26$. Assuming that the availability of the $F_t$ free channels can be uniformly distributed over the total spectrum, the probability of getting $i$, $0 \leq i \leq n$, free channels out of $n$ channels chosen at random follows hypergeometric distribution and is given by $\frac{{F_t \choose i}{C-F_t \choose n-i}}{{C \choose n}}$. The expected number of free channels over all possible situations is then given by $\sum\limits_{i=0}^{n}{\frac{i{F_t \choose i}{C-F_t \choose n-i}}{{C \choose n}}} = \sum\limits_{i=1}^{n}{\frac{F_t{F_t -1 \choose i-1}{C-F_t \choose n-i}}{{C \choose n}}} = \frac{F_t {C-1 \choose n-1}}{{C \choose n}} = nf$, where $f = \frac{F_t}{C}$. Thus, on an average, the number of reserved channels by the first attempt is equal to $nf$. When all channels are free, $f=1$ and all the required $n$ channels are reserved in the first attempt. If $f<1$, then the remaining number of channels to be allocated after the first attempt is $n-nf = n(1-f)$, on an average. For the second attempt, since $F_t - nf$ is the number of free channels, the success probability for getting a free channel will again be a hypergeometric distribution, leading to $nf(1-\frac{n}{C})$ channels reserved by the second attempt on an average. Thus, on an average, after the second attempt, the total number of reserved channels is $min\{n,nf+nf(1-\frac{n}{C})\}$ and the number of channels yet to be allocated is $n-\{nf+nf(1-\frac{n}{C})\}$. Generalizing this observation, we have the following result.

\begin{lemma}\label{lemma1}
The expected number of channels reserved during the $(k+1)^{th}$ attempt, $k \geq 0$, is $nf(1-\frac{kn}{C})$. Also, on an average, the total number of channels reserved after the $k^{th}$ attempt is $min(n, nkf)$.
\end{lemma}

{\bf Proof :} We prove the result by induction. From the discussion above, the proposition that {\it at the $k^{th}$ attempt, the number of channels reserved is $nf\{1-\frac{(k-1)n}{C}\}$ on an average}, is true for $k$ $= 1$ and $2$. Let us assume that this proposition is true for $k = k$.

Hence, at the $(k+1)^{th}$ attempt, the expected number of channels reserved is equal to \\ $\frac{n(F_t- [nf+nf(1-\frac{n}{C}) + \cdots + nf\{1-\frac{(k-1)n}{C}\} ])}{C}$ $\approx n f(1-\frac{kn}{C})$.

Hence, on an average, the total number of channels reserved after the $k^{th}$ attempt is $min[n, \{nf+nf(1-\frac{n}{C}) + nf(1-\frac{2n}{C}) + \cdots + nf(1-\frac{(k-1)n}{C}\}]$ $\approx min(n,nkf)$. \hfill $~\qed$

\begin{theorem}\label{theorem1}
To reserve $n$ channels, the required number of attempts, on an average, is equal to $\lceil\frac{1}{f}\rceil$.
\end{theorem}

{\bf Proof:} By lemma~\ref{lemma1}, the total number of channels reserved after $k^{th}$ attempt is $min(n,nkf)$, on an average. Hence, if $\alpha$ is the minimum number of attempts required for reserving all the $n$ channels, then $n f \alpha \geq n$, i.e., $\alpha \geq \frac{1}{f}$, on an average. Hence the theorem. \hfill $\qed$

\begin{remark}\label{remark1}
Theorem~\ref{theorem1} basically establishes that more the number of free channels, less is the average number of attempts for acquiring the required numbers of channels.
\end{remark}

\begin{corollary}\label{corollary1}
On an average, reservation of all $n$ channels can be done in $\psi = \lceil\frac{1}{f}\rceil \zeta + O(1)$ time, where $\zeta$ is the time for a single execution of the loop in the channel allocation algorithm.
\end{corollary}

It may be noted that, if we use {\em Bit-Map}~\cite{b114, b115} protocol for communication through $CCC$, then $\zeta$ is $O(\Delta)$ time, where $\Delta$ is the maximum node degree of the network as already mentioned in Section~\ref{PIm}.

\subsection{Performance of Channel Allocation Algorithm using $OFDM-FDMA$ Technique}

In order to evaluate the theoretical performance of the Algorithm~\ref{algo4}, let us assume that we try to reserve $n$ channels in successive attempts, where each attempt corresponds to a single execution of steps $6$ to $34$ of the algorithm. We first generate $n$ random numbers $c_k^1$, $1 \leq k \leq n$. Note that $c_k^1$ is one of the $n$ channels chosen at random in which the probability that $i$ channels will be free is given by the hypergeometric distribution as in the case of $FDM-FDMA$ allocation. However, in Algorithm~\ref{algo4}, we also check whether the two adjacent channels $c_k^2 = c_k^1 + 1$ and $c_k^3 = c_k^1 + 2$ are free. There can be eight different possibilities regarding the status of these three channels as depicted in Table \ref{tc1}, where an entry is '$0$' if the corresponding channel is free, and '$1$' if it is blocked. Note that the probability that the channel $c_k^1 + 1$ or $c_k^1 + 2$ will be free, is given by $\frac{F_t}{C} = f$. Table \ref{tc1} shows the probability of selections of $c_k^2$ and $c_k^3$ in each of the eight possible situations with the corresponding number of free channels found.

\begin{table}[]
\caption{{\small All possible cases about the status of three consecutive channels}}\label{tc1}
\centering
\tiny
\begin{tabular}{|c|c|c|l|c|c|c|c|c|c|c|c|c|} \hline
  Status of $c_k^1$ & Status of $c_k^2$ & Status of $c_k^3$ & Probability of selecting $c_k^2$ and $c_k^3$ as free & Number of free channels \\ \hline
  $0$     & $0$     & $0$     & $f^2$     & $3$    \\
  $0$     & $0$     & $1$     & $f.(1-f)$ & $2$    \\
  $0$     & $1$     & $0$     & $f.(1-f)$ & $1$    \\
  $0$     & $1$     & $1$     & $(1-f)^2$ & $1$    \\
  $1$     & $0$     & $0$     & $f^2$     & $2$    \\
  $1$     & $0$     & $1$     & $f.(1-f)$ & $1$    \\
  $1$     & $1$     & $0$     & $f.(1-f)$ & $1$    \\
  $1$     & $1$     & $1$     & $(1-f)^2$ & $0$     \\ \hline
 \end{tabular}
\end{table}

From Table \ref{tc1}, given that the channel $c_k^1$ is free, the expected number of free channels selected out of these three consecutive channels is given by $3f^2 + 2 f (1-f) + f(1-f) + (1-f)^2 = f^2 + f + 1$. Similarly, given that the channel $c_k^1$ is blocked, the expected number of free channels selected out of these three consecutive channels is given by $2f^2 + 2f(1-f) = 2f$. Thus, the expected number of free channels over all possible situations is given by,

$\sum\limits_{i=0}^{n}{i.(f^2+f+1)\frac{{F_t \choose i}{C-F_t \choose n-i}}{{C \choose n}}} + \sum\limits_{i=0}^{n}{(n-i).2f \frac{{F_t \choose i}{C-F_t \choose n-i}}{{C \choose n}}} = n (3f - f^2 + f^3) \approx 3nf$, when $f << 1$.

Thus, in a heavy traffic condition, when $f$ is very small, the average number of attempts to reserve the required number of channels made by Algorithm~\ref{algo4} is approximately equal to $\lceil\frac{1}{3f}\rceil$.

\section{Simulation of Channel Allocation Algorithm}\label{simulation}

In this section, we show the results of simulating our proposed protocol, and evaluate its performance in terms of the average number of attempts made by the proposed algorithm for acquiring the required number of channels to communicate a given multimedia signal and also in term of success rate. We also compare our proposed protocol with first-fit and best-fit channel allocation techniques. Simulations are performed $10,000$ times on random network topologies with each of $100$ to $1100$ nodes, in which nodes are distributed randomly within an area of $(100 \times 100)m^2$. The number of channels required by a node for communication is also varied from $1$ to $8$. We assume that, the signal to be sent has a mix of different multimedia signal types with the proportion of $50\%$, $20\%$, $15\%$, $10\%$ and $5\%$ for voice, data, still image, video and online streaming data, respectively, with demand number of channels ($DN$) as $1,~2,~4,~6$ and $8$, respectively.

For $FDM-FDMA$ technique, the simulation is performed with values of each of the primary and secondary non-overlapping channels as $500$, each, as shown in Fig.~\ref{ONOC}(a). Thus, $C$, the total number of channels is $1000$. The primary channels are assumed to be uniformly distributed over the whole spectrum. We assume that on an average, $30\%$ of the primary channels are used by primary users for different broadcasting purposes. That is, $150$ primary channels are used by different broadcasting purposes and the rest $70\%$ are idle~\cite{b2}, and those $70\%$ will be available for cognitive radio users. For  $OFDM-FDMA$ technique, the simulation is performed with values of each of the primary and secondary overlapping channels as $1000$ each, because the width of one non-overlapping channel is equal to that of two overlapping channels, as shown in Fig.~\ref{ONOC}(b). Thus, $C$, the total number of overlapping channels will be taken as $2000$ for the given total communication bandwidth of $1000$ non-overlapping channels.

\begin{table*}[]
\caption{{\small Average number of blocked channels by 2-distance neighbors with different values of range}}\label{t1}
\centering
\tiny
\begin{tabular}{|c|c|c|c|c|c|c|c|c|c|c|c|c|} \hline
 \multirow{2}{*}{Range in meter} & \multicolumn{11}{|c|}{Number of Blocked Channels} \\ \cline{2-12}
      & 100 Nodes & 200 Nodes & 300 Nodes & 400 Nodes & 500 Nodes & 600 Nodes & 700 Nodes & 800 Nodes & 900 Nodes & 1000 Nodes & 1100 Nodes \\ \hline
 \hline
 $10$ & $9$       & $32$      & $55$      & $79$      & $104$     & $129$     & $153$     & $178$     & $203$     & $229$      & $254$      \\ \hline
 $15$ & $33$      & $82$      & $133$     & $184$     & $237$     & $288$     & $340$     & $392$     & $444$     & $497$      & $549$      \\ \hline
 $20$ & $62$      & $143$     & $227$     & $312$     & $397$     & $481$     & $565$     & $652$     & $736$     & $819$      & $905$      \\ \hline
 $25$ & $96$      & $212$     & $334$     & $453$     & $571$     & $689$     & $811$     & $927$     & $1048$    & $1166$     & $1283$     \\ \hline
 \end{tabular}
\end{table*}

When a cognitive radio user wants to transmit a multimedia signal, the average number of channels blocked by all of its neighbors up to distance two, with different values of range are shown in Table~\ref{t1}. Note that these values exclude the $150$ primary channels used for various broadcasting purposes. From Table~\ref{t1}, the number of blocked channels by all 2-distance neighbors increases with the sensing and transmission range as expected.

\begin{table}[]
\caption{{\small Average number of free channels for $500$ and $700$ nodes in Non-Overlapping Channel}}\label{t2}
\centering
\scriptsize
\begin{tabular}{|c|c|c|c|} \hline
\parbox{1.25in}{Average number of blocked channels} &
\parbox{1.25in}{Average number of free primary channels ($F_P$)} &
\parbox{1.25in}{Average number of free secondary channels ($F_S$)}&
\parbox{1.25in}{Average number of free channels ($F$)} \\ \hline
 \hline
 \multicolumn{4}{|c|}{{\bf For $500$ nodes}} \\ \hline
 \hline
  $254$ & $307$ & $439$ & $746$ \\ \hline
  $387$ & $252$ & $361$ & $613$ \\ \hline
  $547$ & $187$ & $266$ & $453$ \\ \hline
  $721$ & $115$ & $164$ & $279$ \\ \hline
\hline
 \multicolumn{4}{|c|}{{\bf For $700$ nodes}}\\ \hline
 \hline
  $303$ & $287$ & $410$ & $697$ \\ \hline
  $490$ & $210$ & $300$ & $510$ \\ \hline
  $715$ & $117$ & $168$ & $285$ \\ \hline
  $961$ & $16$  & $23$  & $39$  \\ \hline
\end{tabular}
\end{table}

\begin{table}[]
\caption{{\small Average number of free channels for $700$ and $1000$ nodes in Overlapping Channel}}\label{t2_1}
\centering
\scriptsize
\begin{tabular}{ |c|c|c|c| } \hline
\parbox{1.25in}{Average number of blocked channels} &
\parbox{1.25in}{Average number of free primary channels ($F_P$)} &
\parbox{1.25in}{Average number of free secondary channels ($F_S$)}&
\parbox{1.25in}{Average number of free channels ($F$)} \\ \hline
 \hline
 \multicolumn{4}{|c|}{{\bf For $700$ nodes}} \\ \hline
 \hline
  $303$ & $545$ & $640$ & $1185$  \\ \hline
  $490$ & $367$ & $431$ & $798$ \\ \hline
  $715$ & $210$ & $246$ & $456$ \\ \hline
  $961$ & $97$  & $114$ & $211$ \\ \hline
 \hline
 \multicolumn{4}{|c|}{{\bf For $1000$ nodes}}\\ \hline
 \hline
  $379$  & $466$ & $549$ & $1015$ \\ \hline
  $647$  & $251$ & $295$ & $546$  \\ \hline
  $969$  & $94$  & $111$ & $205$  \\ \hline
  $1316$ & $10$  & $11$  & $21$     \\ \hline
\end{tabular}
\end{table}

For $FDM-FDMA$ channel allocation technique, Table~\ref{t2} shows the values of $F_s$, $F_p$ and $F = F_p + F_s$ for $500$ and $700$ nodes, respectively for different values of range used in the simulation experiment. Column $1$ of Table~\ref{t2} also show the total number of blocked channels, i.e., the channels allocated to all nodes up to 2-distance neighbors of a transmitting node for avoiding the hidden node problem (which are actually taken from Table~\ref{t1} corresponding to $500$ and $700$ nodes, respectively), plus $150$ broadcasting channels blocked by primary users. Table~\ref{t2_1}, for  $OFDM-FDMA$ channel allocation technique, show the values of $F_s$, $F_p$ and $F = F_p + F_s$ for $700$ and $1000$ nodes, respectively for different values of range used in the simulation experiment. Column $1$ of Table~\ref{t2_1} also show the total number of blocked channels, i.e., the channels allocated to all nodes up to 2-distance neighbors of a transmitting node for avoiding the hidden node problem (which are actually taken from Table~\ref{t1} corresponding to $700$ and $1000$ nodes, respectively). We notice from Table~\ref{t2_1} that the total number of free channels $F$ is much less than the difference between the total number of channels (which is $1000$ in our case) and the number of blocked channels given in column $1$ of these corresponding Table~\ref{t2_1}. This is because of the fact that for avoiding channel interferences between two distinct pairs of communicating nodes, we maintain a gap of one channel on either side of a channel allocated to a communicating pair of nodes for computing the number of free channels to be allocated to other users. From Table~\ref{t2_1}, we see that with $700$ nodes, the number of free channels can go down to $211$, i.e., $10.55\%$ of the total number of channels for a range of $25$ meters. This situation corresponds to a moderate traffic in the network. On the other hand, with $1000$ nodes and a range of $25$ meters, the number of free channels can go down to as low as only $21$, which corresponds to a very heavy traffic in the network.

The average number of attempts for different values of $DN$ (corresponding to different multimedia signal types) with different percentage values of blocked channels for $500$ and $700$ nodes for $FDM-FDMA$ channel allocation technique and $700$ and $1000$ nodes for  $OFDM-FDMA$ channel allocation technique are shown in Tables~\ref{t3} and~\ref{t3_1}, respectively. The values in these tables capture the behavior of our proposed algorithm under different traffic load ({\em by load we mean the percentage of blocked channels}) ranging from $30\%$ (light load) to more than $95\%$ (extremely heavy load). Both the Tables~\ref{t3} and~\ref{t3_1} show that when the number of free channels decreases, the average number of attempts increases, as expected. For $FDM-FDMA$ technique, when the number of blocked channels lies between $30\%$ and $70\%$, we require only $2$ to $4$ attempts. However, in the most unlikely situations of an extremely heavy traffic load with about $96\%$ blocked channels, the average number of attempts will be $29$ for $DN=8$, as shown in Table~\ref{t3}. For $OFDM-FDMA$ technique, when the number of blocked channels lies between $50\%$ and $70\%$, the average number of attempt will be equal to $2$, while with about $90\%$ blocked channels, the algorithm needs at most $4$ attempts on an average. For extremely heavy traffic load with about $99\%$ blocked channels, the average number of attempt is equal to $45$ for $DN = 8$, as shown in Table~\ref{t3_1}. We, however, note that the sensing time per attempts in  $OFDM-FDMA$ channel allocation technique is three times more than that in $FDM-FDMA$ channel allocation technique, as already mentioned in Section~\ref{SAC1}. We see from Tables~\ref{t3} and~\ref{t3_1} that the number of attempts found through simulation agrees well with the theoretical values except when the traffic is extremely heavy, e.g., $96\%$ blocked channels for $FDM-FDMA$ technique and $99\%$ blocked channels for $OFDM-FDMA$ technique. This small deviation may arise due to randomness in the simulation process.

\begin{figure}[]
  \centering
  \includegraphics[width=3.0in,height=2.25in]{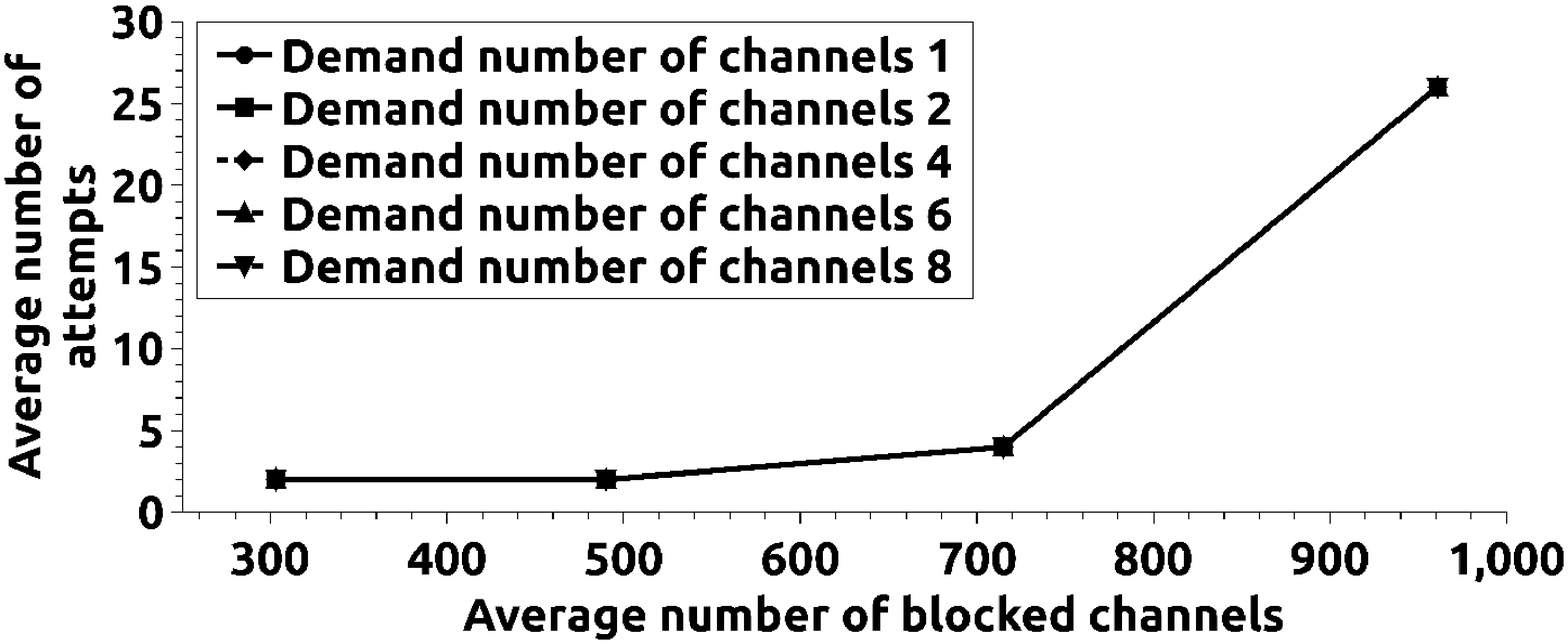}
  \caption{Number of attempts vs percentage of blocked channel for $700$ nodes with $FDM-FDMA$}\label{fpbvsna}
\end{figure}

\begin{figure}[]  
  \centering
  \includegraphics[width=3.0in,height=2.25in]{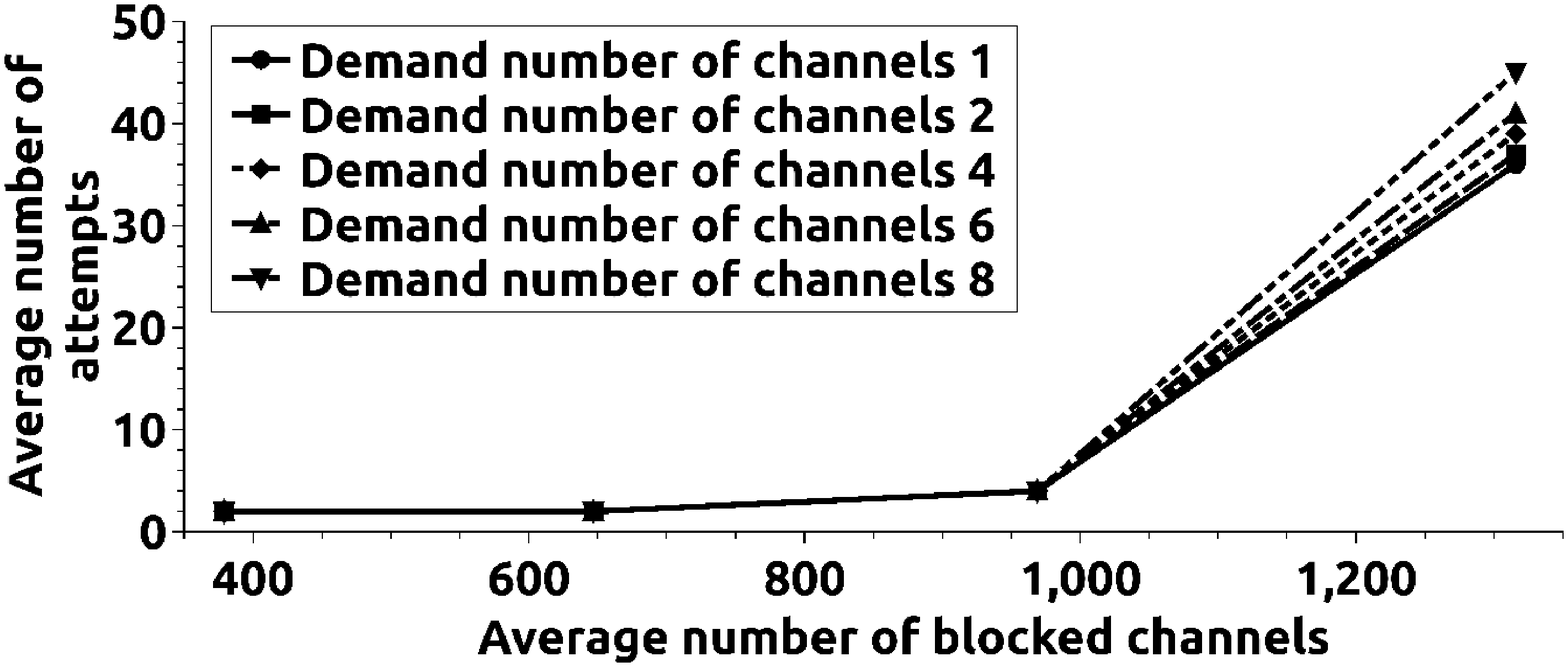}
  \caption{Number of attempts vs percentage of blocked channel for $1000$ nodes with $OFDM-FDMA$}\label{fpbvsna1}
\end{figure}

Tables~\ref{t3} and~\ref{t3_1} also show the performance comparison of the proposed protocol with the first-fit and best-fit allocation techniques. We see that our proposed protocol is superior to either of them under all traffic situations in respect of average number of attempts as well as the success rate, {\em where the success rate is defined as the percentage of the cases the protocol in question can successfully allocate channels}. Thus, with $DN = 8$ and traffic load of $70\%$, our protocol with $FDM-FDMA$ technique requires only $4$ attempts as opposed to $500$ attempts using first-fit and $638$ attempts using best-fit allocations. Similarly, with $DN = 8$ and traffic load of $70\%$, our protocol with $OFDM-FDMA$ technique requires only $2$ attempts as opposed to $250$ attempts using first-fit and $639$ attempts using best-fit allocations. Further, with $DN = 8$, using $FDM-FDMA$ technique, the success rate of our protocol is always $100\%$ even up to a traffic load of $96\%$, while both the first-fit and best-fit techniques fail to allocate any channel under this condition. Similarly, with $DN = 8$ and a traffic load of $99\%$, using $OFDM-FDMA$ technique, neither of the first-fit and best-fit techniques can allocate the channels, although our proposed protocol has the success rate of $100\%$. The nature of variation of the simulated values of the required number of attempts under different traffic conditions with $700$ (for $FDM-FDMA$) and $1000$ (for $OFDM-FDMA$) nodes is also shown in Fig.~\ref{fpbvsna} and~\ref{fpbvsna1}, respectively for different values of $DN$.

\begin{table*}[]
\caption{{\small Comparison of $FDM-FDMA$ and $OFDM-FDMA$ for different types of multimedia signal with different nodes and range $25$ meter}}\label{ttx}
\centering
\scriptsize
 \begin{tabular}{|c|c|c|c|c|c|c|c|c|c|c|c|} \hline

 \multirow{3}{*}{\parbox{1cm}{$~$ \\Number of nodes\\}} & \multicolumn{5}{|c|}{Success rate in $FDM-FDMA$}      & \multicolumn{5}{|c|}{Success rate in $OFDM-FDMA$}        \\ \cline{2-11}
                                                        & \multicolumn{5}{|c|}{Demand Number of channels} & \multicolumn{5}{|c|}{Demand Number of channels}    \\ \cline{2-11}
                                                        & $8$   & $6$   & $4$   & $2$   & $1$              & $8$    & $6$    & $4$    & $2$   & $1$              \\ \hline
  $100$ nodes                                           & $100$ & $100$ & $100$ & $100$ & $100$            & $100$  & $100$  & $100$  & $100$ & $100$            \\ \hline
  $200$ nodes                                           & $100$ & $100$ & $100$ & $100$ & $100$            & $100$  & $100$  & $100$  & $100$ & $100$            \\ \hline
  $300$ nodes                                           & $100$ & $100$ & $100$ & $100$ & $100$            & $100$  & $100$  & $100$  & $100$ & $100$            \\ \hline
  $400$ nodes                                           & $100$ & $100$ & $100$ & $100$ & $100$            & $100$  & $100$  & $100$  & $100$ & $100$            \\ \hline
  $500$ nodes                                           & $100$ & $100$ & $100$ & $100$ & $100$            & $100$  & $100$  & $100$  & $100$ & $100$            \\ \hline
  $600$ nodes                                           & $100$ & $100$ & $100$ & $100$ & $100$            & $100$  & $100$  & $100$  & $100$ & $100$            \\ \hline
  $700$ nodes                                           & $100$ & $100$ & $100$ & $100$ & $100$            & $100$  & $100$  & $100$  & $100$ & $100$            \\ \hline
  $800$ nodes                                           & $0$   & $0$   & $0$   & $0$   & $0$              & $100$  & $100$  & $100$  & $100$ & $100$            \\ \hline
  $900$ nodes                                           & $0$   & $0$   & $0$   & $0$   & $0$              & $100$  & $100$  & $100$  & $100$ & $100$            \\ \hline
  $1000$ nodes                                          & $0$   & $0$   & $0$   & $0$   & $0$              & $100$  & $100$  & $100$  & $100$ & $100$            \\ \hline
  $1100$ nodes                                          & $0$   & $0$   & $0$   & $0$   & $0$              & $0$    & $0$    & $0$    & $0$   & $0$              \\ \hline
 \end{tabular}
\end{table*}

\begin{figure}[]  
  \centering
  \includegraphics[width=3.0in,height=2.25in]{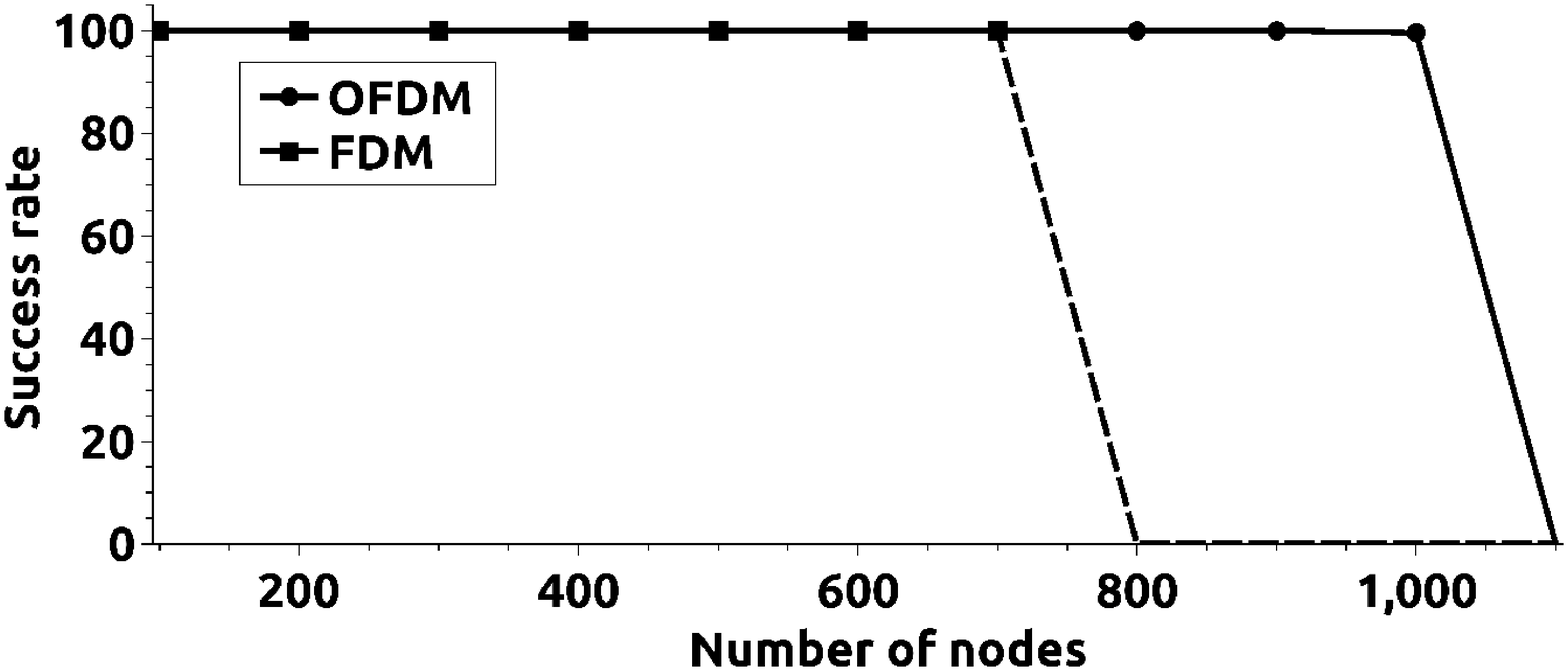}
  \caption{Success rate comparison for $FDM-FDMA$ vs $OFDM-FDMA$ with $8$ demand number of channels}\label{FDMvsOFDM}
\end{figure}

Table~\ref{ttx} and Fig.~\ref{FDMvsOFDM} show the performance comparison between the $FDM-FDMA$ and $OFDM-FDMA$ implementations with our protocol. From Table~\ref{ttx} and Fig.~\ref{FDMvsOFDM}, we notice that with a range of $25$ meter, $FDM-FDMA$ can not allocate the channel beyond $700$ nodes, while $OFDM-FDMA$ can work satisfactorily even with $1000$ nodes.

\subsection{Execution Time of the Proposed Protocol}\label{overhead}

Assuming that the address fields follow $IPv4$ (or $IPv6$) format, the maximum length of a message for channel sensing (e.g., $TAM$) will be $10$ (or $34$) bytes which need $1.25$ (or $4$) msec with a bandwidth of $64$ Kbps for $CCC$. From the data given in Table~\ref{t3}, we see that for video transmission with about $70\%$ traffic load through eight $64$ Kbps channels, four attempts are needed by the Algorithm~\ref{algo3} with $FDM-FDMA$ technique, which corresponds to an overhead of $4 \times 8 \times 1.25$ msec $= 40$ msec with $IPv4$ format ($4 \times 8 \times 4 = 128$ msec with $IPv6$ format). On the other hand, with $OFDM-FDMA$ and about $70\%$ blocked channels, Algorithm~\ref{algo4} makes $2$ attempts, leading to $2 \times 3 \times 8 \times 1.25 = 60$ msec with $IPv4$ format ($2 \times 3 \times 8 \times 4 = 192$ msec with $IPv6$ format) for $DN = 8$.

\subsection{Delay and Delay Jitter in Transmission}\label{addj}

In this section, we show our experimental results through simulation by using our proposed scheme on different types of multimedia data. The file categories and the test files have been chosen so as to reflect as closely as possible the different types of wireless data communication that can be seen in today's world. For this purpose, we choose files categorized into four different file types, namely, video, music, image and text. Below we provide a description of files chosen from each of these categories :

\begin{enumerate}
\item {\bf Video files} : The video files are in $MPEG$ and $DAT$ format \cite{video}. Average size of these files is about $48$ $MB$.
\item {\bf Music files} : The music files are in $MP3$ encoded format \cite{music}. Average size of these files is about $5.87$ $MB$.
\item {\bf Image files} : The image files are in $JPEG$ format \cite{image}. Average size of these files is about $1.5$ $MB$.
\item {\bf Text files} : The text files are in $TXT$ file format \cite{data}. Average size of these files is about $80.89$ $KB$.
\end{enumerate}

For our experiment with $FDM-FDMA$ technique, we assume a data frame size of $1024$ bytes using $IPv4$ packet format. We also assume that the channel request and release rates of a user are $0.7$ and $0.3$, respectively. Each file was tested $1000$ times during the simulation to obtain the average delay and delay jitter as reported in the Tables~\ref{different_video_file}, \ref{different_music_file}, \ref{different_image_file} and \ref{different_data_file}.

In each of the Tables~\ref{different_video_file}, \ref{different_music_file}, \ref{different_image_file} and \ref{different_data_file}, the first column represents the file size. For video transmission we use $8$ channels and for music, images and data we use $6$, $4$ and $2$ channels, respectively. The next column represents the average number of channels deallocated by $PU$s when a $PU$ asks for its channel currently used by the $SU$. The third column represents ideal transmission ($IT$) time for transmitting files (with zero overhead). The next column represents initial channel allocation ($ICA$) time for grabbing all the required $DN$ channels before starting the communication. The fifth column represents channel reallocation ($CR$) time when a $SU$ has to deallocate a channel and then a new channel is re-allocated to the $SU$. The, actual transmission ($AT$) time is equal to $IT + ICA + CR$, which is given in the next column. The seventh column represents the maximum possible jitter between two consecutive packets due to channel reallocation. The next column represents the mean jitter and the last column represents the standard deviation of jitter. The appropriate units for the values in different columns have been mentioned in the tables. In each of the above tables, the bold-faced entries in a row represent the average behavior for the corresponding traffic load.

These simulation results show that the overhead in time due to the proposed allocation algorithm constitutes a very small fraction of the total transmission time. For example, from Table~\ref{different_video_file}, we see that with a traffic load of about $70\%$, for transmitting video files of size about $48$ MB, the total transmission time through a $64$ Kbps channel is around $786.5$ sec, while the overhead due to channel allocation and reallocation required by our proposed algorithm is approximately $49$ msec ($\approx 0.0062 \%$). 
  
During the transmission of the packets, a delay may appear between the transmission of two consecutive data packets whenever a $PU$ channel is allocated and that channel is to be released due to the demand from the corresponding $PU$. This delay is due to the execution of our proposed algorithm for reallocating the channels which varies from zero to a finite amount of time, causing delay jitter. We have estimated this delay jitter for all the above cases of our simulation experiment with different types of real-life multimedia data and find that this jitter is very small. For example, from Table~\ref{different_video_file}, we see that with a traffic load of about $70\%$, for transmitting video files, the mean jitter is around $0.0015$ msec with a standard deviation of about $0.064$ msec, whereas the transmission time of a sub-packet of size $1024$ bytes through $64$ Kbps channel is $125$ msec.

\section{Analysis of Protocol by Markov Model}

To analyze the performance of the proposed protocol for channel allocation and deallocation, we model the system by the Markov's birth and death process, where channel allocation corresponds to the birth process and deallocation corresponds to the death process. At any time instant, the allocation status of the system can be designated by a state $S_{k}^{k'}$ of the system, where $k$ represents the number of reserved secondary channels and $k'$ represents that of the reserved primary channels. The system will start transmitting the message when all the $n$ required channels are reserved, i.e., $k+k'=n$, and then allocated by the Algorithm~\ref{algo3}. At any state $S_{k}^{k'}$, when a single new (free) channel is reserved by the Algorithm~\ref{algo3}, the system can move either to the state $S_{k}^{k'+1}$ (if the new channel is reserved from the primary band) or to the state $S_{k+1}^{k'}$ (if the new channel is reserved from the secondary band). Similarly, if $i$ channels are reserved in one attempt (Algorithm~\ref{algo3}), out of which $i_1$ channels ($0 \le i_1 \le i$), are in the primary band and the rest $i-i_1$ are in the secondary band, then the system will move from the state $S_{k}^{k'}$ to the state $S_{k+i-i_1}^{k'+i_1}$.

Let $\mu_1 $ be the probability per unit time for reserving one new channel in one attempt (either from the primary band or from the secondary band). Then the probability per unit time that the system moves from the state $S_{k}^{k'}$ to $S_{k}^{k'+1}$ is given by $\frac{{F_{p,t} \choose 1}}{{F_t \choose 1}}\mu_1$ and the probability per unit time that the system moves from the state $S_{k}^{k'}$ to $S_{k+1}^{k'}$ is given by $\frac{{F_{s,t} \choose 1}}{{F_t \choose 1}}\mu_1$. In a similar manner, let $\mu_2 $ be the probability per unit time for reserving two new channels in one single attempt. Then the system can move from the state $S_{k}^{k'}$ to $3$ possible states $-$ $i)$ to the state $S_{k}^{k'+2}$ with a probability of $\frac{{F_{p,t} \choose 2}}{{F_t \choose 2}}\mu_2$, $ii)$ to the state $S_{k+2}^{k'}$ with a probability of $\frac{{F_{s,t} \choose 2}}{{F_t \choose 2}}\mu_2$, and $iii)$ to the state $S_{k+1}^{k'+1}$ with a probability of $\frac{{F_{p,t} \choose 1}{F_{s,t} \choose 1}}{{F_t \choose 2}}\mu_2$. Similarly, when $i$ channels are reserved in one single attempt, the transition probabilities from the state $S_{k}^{k'}$ to $i+1$ different possible states can be expressed in terms of $\mu_i$, the total probability per unit time for reserving $i$ channels, $F_{p,t}, F_{s,t}$ and $F_t$. We assume that the time required for allocating the required number $n$ of channels is small enough so that the values of $F_{p,t}, F_{s,t}$ and $F_t$ do not change during the allocation process. Thus, in all the expressions above for transition probabilities, we replace $F_{p,t}, F_{s,t}$ and $F_t$ by time-invariant values $F_p, F_s$ and $F$, respectively. Let $\lambda$ be the probability per unit time that a grabbed primary channel is released by the system due to the arrival of a channel allocation request from a primary user. For the time being, we assume that only one primary channel may be released at any instant of time. Also, let $T$ be the total time required for completing the communication process of a multimedia message. That means, $\frac{1}{T}$ is the probability per unit time that the system moves from the state $S_{k}^{n-k}$, $0 \leq k \leq n$, to the state $S_{0}^{0}$. Further, let $\sigma$ be the probability per unit time that the Algorithm~\ref{algo3} terminates unsuccessfully (when all the requested number of channels could not be allocated) after the time-out period.

\subsection{1-Channel System}

\begin{figure}[]
\centering
  \includegraphics[width=1.75in,height=1.0in]{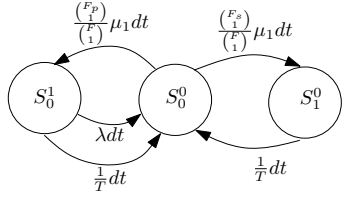}
\caption{State transition diagram for 1-channel system}\label{f23}
\end{figure}

Suppose we want to transmit a voice signal for which $n = 1$. In this case, the system will have only three possible states $-$ $i)$ $S_0^0$, when no channel has been reserved, $ii)$ $S_0^1$, at which one channel is reserved from the primary band and $iii)$ $S_1^0$, at which one channel is reserved from the secondary band. Let $p_i^j, (0 \le i,j \le 1)$ be the probability that the system is in state $S_i^j$. The state transition diagram for this system is as shown in Fig.~\ref{f23}. The different transition arcs are labeled with the corresponding probabilities of transition during a small time interval $dt$. Using the principle of detailed balance for transitions between the states $S_0^0$ and $S_0^1$, we can write,
\begin{equation}\label{e1}
\left(\lambda + \frac{1}{T}\right) p_0^1 = \frac{{F_{p} \choose 1}}{{F \choose 1}} \mu_1 p_0^0
\end{equation}
Similarly, for the transitions between the states $S_0^0$ and $S_1^0$, we have,
\begin{equation}\label{e2}
\frac{1}{T} p_1^0 = \frac{{F_{s} \choose 1}}{{F \choose 1}} \mu_1 p_0^0
\end{equation}
Since the system must be in one of the three states, we have,
\begin{equation}\label{e3}
p_0^0 + p_1^0 + p_0^1 = 1
\end{equation}
Substituting the values of $p_0^1$ and $p_1^0$ from eqns.~(\ref{e1}) and (\ref{e2}) in eqn.~(\ref{e3}), we get,
\begin{equation}
p_0^0 + \left( \frac{\frac{{F_{p} \choose 1}}{{F \choose 1}}}{\lambda + \frac{1}{T}} \right) \mu_1 p_0^0 + \left( \frac{\frac{{F_{s} \choose 1}}{{F\choose 1}}}{\frac{1}{T}} \right) \mu_1 p_0^0 = 1
\end{equation}
That is,
\begin{equation}\label{e4}
p_0^0 = \frac{1}{1 + \left( \frac{\frac{{F_{p} \choose 1}}{{F \choose 1}}}{\lambda + \frac{1}{T}} \right) \mu_1 + \left(
\frac{\frac{{F_{s} \choose 1}}{{F \choose 1}}}{\frac{1}{T}} \right) \mu_1 }
\end{equation}
We say that the system in the {\em active} condition when the required numbers of channels are reserved, and the probability for the system being in that condition is given by,
\begin{equation}\label{e5}
P_1 = p_1^0 + p_0^1 = \frac{ \left( \frac{\frac{{F_{p} \choose 1}}{{F \choose 1}}}{\lambda + \frac{1}{T}} \right) \mu_1 + \left( \frac{\frac{{F_{s}
\choose 1}}{{F \choose 1}}}{\frac{1}{T}} \right) \mu_1 }{1 + \left( \frac{\frac{{F_{p} \choose 1}}{{F \choose 1}}}{\lambda + \frac{1}{T}} \right)
\mu_1 + \left( \frac{\frac{{F_{s} \choose 1}}{{F \choose 1}}}{\frac{1}{T}} \right) \mu_1 }
\end{equation}

We already assume that $T$ is the total time required for completing the communication process of a multimedia message. We also find that the probability for the system being in active state $P_1$. So, the average time for communicating the multimedia data is become $\frac{T}{P_1}$. Thus, the average waiting time can be expressed as $\Gamma_1 = T(\frac{1-P_1}{P_1})$.

\subsection{2-Channel System}

\begin{figure}[]
\centering
  \includegraphics[width=3.5in,height=3.0in]{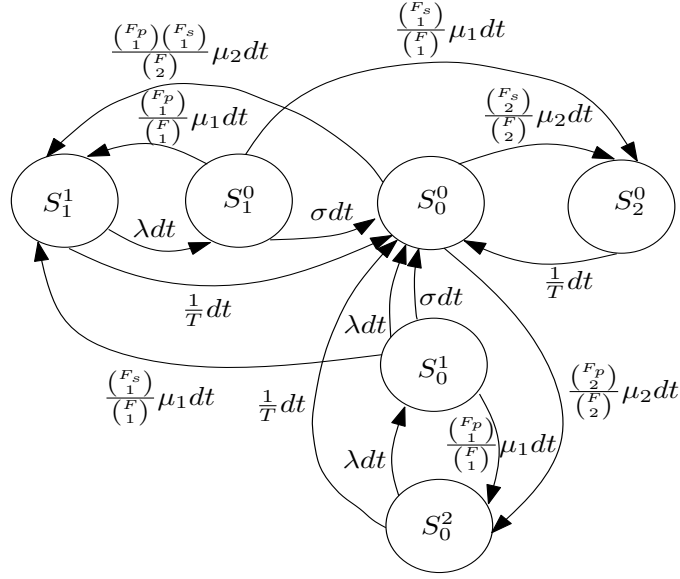}
\caption{State transition diagram for 2-channel system}\label{f22}
\end{figure}

We now consider the case for the 2-channel system for which $n = 2$. As in the case of 1-channel system, we draw the state transition diagram for this system as shown in Fig.~\ref{f22}, consisting of the six states $S_0^0,S_0^1,S_1^0, S_1^1, S_0^2$ and $S_2^0$.

Considering all possible transitions to and from the state $S_2^0$, we have,
\begin{equation}\label{e21}
\frac{1}{T} p_2^0 = \frac{{F_{s} \choose 2}}{{F \choose 2}} \mu_2 p_0^0 + \frac{{F_{s} \choose 1}}{{F \choose 1}} \mu_1 p_1^0
\end{equation}

Similarly, corresponding to all transitions to and from the state $S_0^2$, we have,
\begin{equation}\label{e22}
\left(\lambda + \frac{1}{T}\right) p_0^2 = \frac{{F_{p} \choose 2}}{{F \choose 2}} \mu_2 p_0^0 + \frac{{F_{p} \choose 1}}{{F \choose 1}} \mu_1 p_0^1
\end{equation}

Corresponding to all possible transitions to and from the state $S_1^1$, we have,
\begin{equation}\label{e23}
\left(\lambda + \frac{1}{T}\right) p_1^1 = \frac{{F_{p} \choose 1}}{{F \choose 1}} \mu_1 p_1^0 + \frac{{F_{s} \choose 1}}{{F \choose 1}} \mu_1 p_0^1 +\frac{{F_{p} \choose 1} {F_{s} \choose 1}}{{F \choose 2}} \mu_2 p_0^0
\end{equation}

Considering all possible transitions to and from the state $S_0^1$, we get,
\begin{equation}\label{e24}
\left\{\lambda + \sigma + \frac{{F_{p}\choose 1}}{{F \choose 1}}\mu_1 + \frac{{F_{s}\choose 1}}{{F \choose 1}}\mu_1 \right\} p_0^1 = \lambda p_0^2
\end{equation}

Similarly, considering all possible transitions to and from the state $S_1^0$, we get,
\begin{equation}\label{e25}
\left\{\sigma + \frac{{F_{p}\choose 1}}{{F \choose 1}}\mu_1 + \frac{{F_{s}\choose 1}}{{F \choose 1}}\mu_1 \right\} p_1^0 = \lambda p_1^1
\end{equation}

We also have the following condition to be satisfied :
\begin{equation}\label{e27}
p_0^0 + p_1^0 + p_0^1 + p_2^0 + p_0^2 + p_1^1 = 1
\end{equation}

The total probability for the system to be in the {\em active} condition, is given by
\begin{equation}\label{e28}
P_2 = p_2^0 + p_0^2 + p_1^1
\end{equation}

Like {\em 1-channel system} we can get the value of $P_2$ and $\Gamma_2$ in terms of $\lambda, ~\sigma, ~\mu, ~T, ~F_s, ~F_p$ and $~F$ easily.

\subsection{3-Channel System}

\begin{figure*}[]
\centering
  \includegraphics[width=5.5in,height=4.5in]{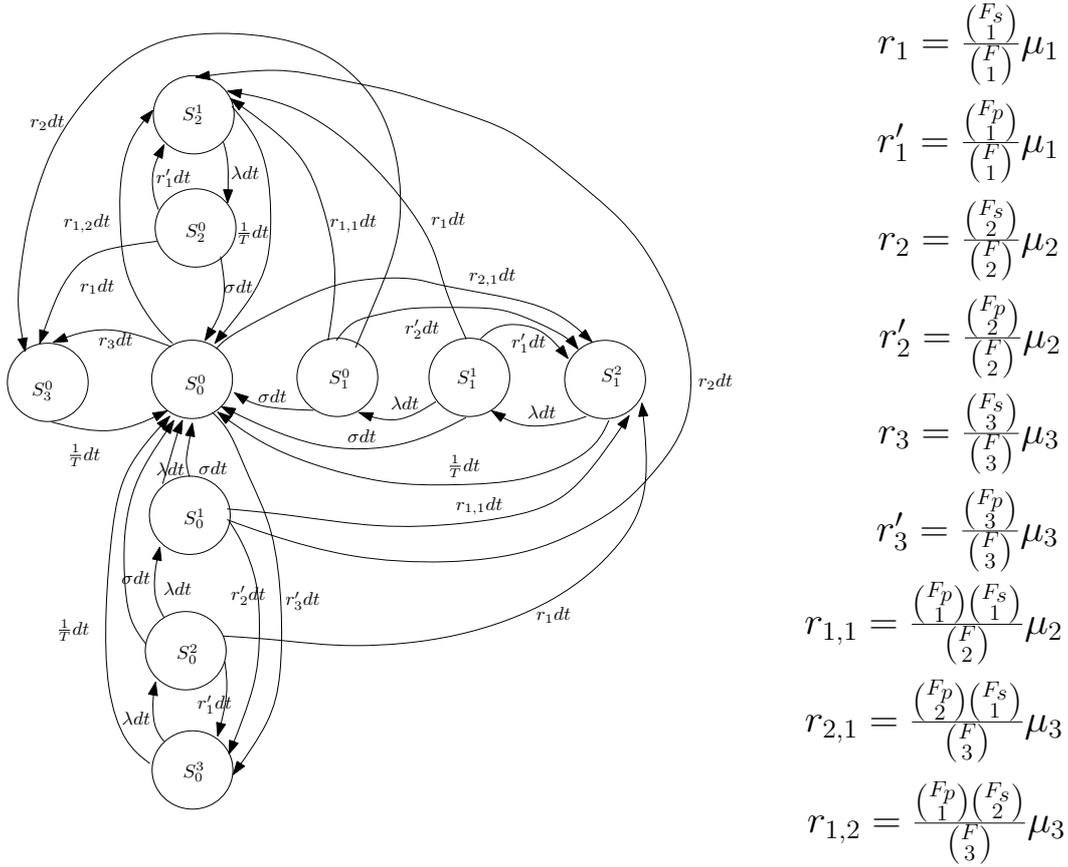}
\caption{State transition diagram for 3-channel system}\label{f24}
\end{figure*}

We now consider the case for the 3-channel system for which $n = 3$. As in the case of 1-channel and 2-channel system, we draw the state transition diagram for this system as shown in Fig.~\ref{f24}, consisting of the ten states $S_0^0$, $S_0^1$, $S_1^0$, $S_1^1$, $S_0^2$, $S_2^0$, $S_2^1$, $S_1^2$, $S_3^0$ and $S_0^3$. The value of $r_1$, $r'_1$, $r_2$, $r'_2$, $r_3$, $r'_3$, $r_{1,1}$, $r_{1,2}$ and $r_{2,1}$ used in the discussion below refer to those given in Fig.~\ref{f24}.

Considering all possible transitions to and from the state $S_2^0$, we have,
\begin{equation}
\left\{\sigma + r'_1 + r_1 \right\} p_2^0 = \lambda p_2^1
\end{equation}

Similarly, considering all possible transitions to and from the state $S_0^2$, we have,
\begin{equation}
\left\{\sigma + \lambda + r'_1 + r_1 \right\} p_0^2 = \lambda p_0^3
\end{equation}

Considering all possible transitions to and from the state $S_1^1$, we get,
\begin{equation}
\left\{\sigma + \lambda + r'_1 + r_1 \right\} p_1^1 = \lambda p_1^2
\end{equation}

Considering all possible transitions to and from the state $S_0^1$, we have,
\begin{equation}
\left\{\sigma + \lambda + r'_2 + r_2 + r_{1,1} \right\} p_0^1 = \lambda p_0^2
\end{equation}

Considering all possible transitions to and from the state $S_1^0$, we get,
\begin{equation}
\left\{\sigma + r'_2 + r_2 + r_{1,1} \right\} p_1^0 = \lambda p_1^1
\end{equation}

Considering all possible transitions to and from the state $S_3^0$, we have,
\begin{equation}
\frac{1}{T}p_3^0 = ( r_3 p_0^0 + r_1 p_2^0 + r_2 p_1^0 )
\end{equation}

Considering all possible transitions to and from the state $S_0^3$, we get,
\begin{equation}
\left(\lambda + \frac{1}{T} \right) p_0^3 = ( r'_3 p_0^0 + r'_2 p_0^1 + r'_1 p_0^2 )
\end{equation}

Considering all possible transitions to and from the state $S_2^1$, we have,
\begin{equation}
\left(\lambda + \frac{1}{T} \right) p_2^1 = ( r_{1,2} p_0^0 + r'_1 p_2^0 + r_2 p_0^1 + r_1 p_1^1 + r_{1,1} p_1^0 )
\end{equation}

Considering all possible transitions to and from the state $S_1^2$, we get,
\begin{equation}
\left(\lambda + \frac{1}{T}\right) p_1^2 = ( r'_1 p_1^1 + r'_2 p_1^0 + r_{2,1} p_0^0 + r_{1,1} p_0^1 + r_1 p_0^2 )
\end{equation}

We also have the following condition to be satisfied :
\begin{equation}
p_0^0 + p_1^0 + p_0^1 + p_2^0 + p_0^2 + p_1^1 + p_3^0 + p_0^3 + p_2^1 + p_1^2 = 1
\end{equation}

The total probability for the system to be in the {\em active} condition, is given by
\begin{equation}
P_3 = p_3^0 + p_0^3 + p_2^1 + p_1^2
\end{equation}

Like {\em 1-channel system} we can get the value of $P_3$ in terms of $\lambda, ~\sigma, ~\mu, ~T, ~F_s, ~F_p$ and $~F$ easily.

\subsection{Examples Showing the Results from Markov Model}

\begin{table}[]
\caption{{\small $P_n$ and $\Gamma_n$ with different message lengths, $F_s=23$ and $F_p=16$ for non-overlapped channels with $700$ nodes}}\label{t4}
\centering
\scriptsize
\begin{tabular}{|c|c|l|l|l|l|l|}
\hline
\multicolumn{2}{|c|}{\multirow{2}{*}{$P_n / \Gamma_n$}} & \multicolumn{5}{|c|}{{$1 / T$}}  \\ \cline{3-7}

\multicolumn{2}{|c|}{} & $0.01$ & $0.25$ & $0.5$ & $0.75$ & $0.99$ \\ \hline
\hline
\multirow{3}{*}{\parbox{1.5cm}{Active State}} & $P_1$ & $0.9769$             & $0.6849$             & $0.5423$          & $0.4517$           & $0.3901$ \\ \cline{2-7}
                                              & $P_2$ & $0.969$              & $0.6512$             & $0.5172$          & $0.4332$           & $0.3758$ \\ \cline{2-7}
                                              & $P_3$ & $0.962$              & $0.6282$             & $0.5012$          & $0.4217$           & $0.367$  \\ \hline
\hline
\multirow{3}{*}{\parbox{1.5cm}{Average Waiting Time (in units of $T$)}} & $\Gamma_1$ & $2.3692$ & $1.8404$ & $1.6883$ & $1.6183$ & $1.5792$ \\ \cline{2-7}
                                                                         & $\Gamma_2$ & $3.1979$ & $2.1426$ & $1.867$  & $1.7446$ & $1.6778$ \\ \cline{2-7}
                                                                         & $\Gamma_3$ & $3.9501$ & $2.3671$ & $1.9903$ & $1.8287$ & $1.7419$ \\ \hline
\end{tabular}
\end{table}

\begin{figure}[]
  \centering
  \includegraphics[width=3.0in,height=2.25in]{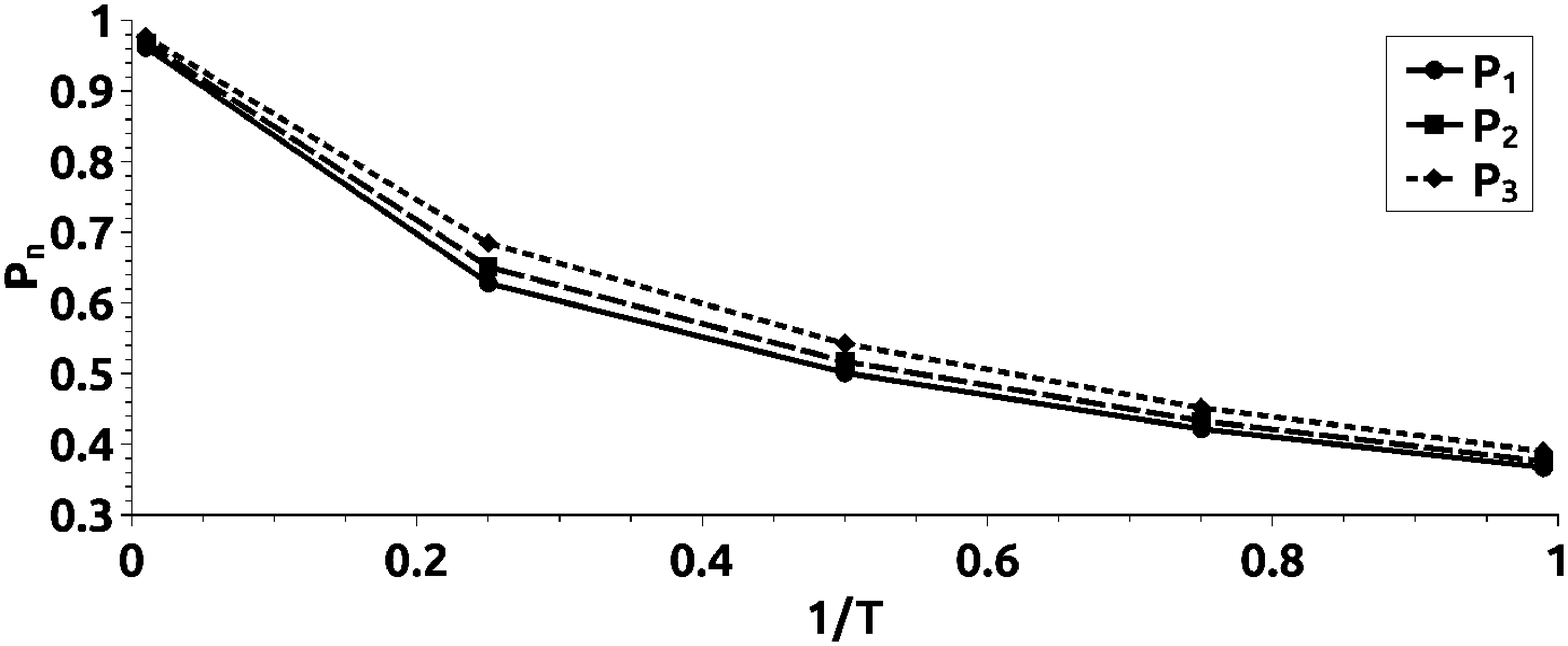}
  \caption{$P_n$ with different lengths of messages for $F_s=23$ and $F_p=16$}\label{Pn1}
\end{figure}

\begin{figure}[]
  \centering
  \includegraphics[width=3.0in,height=2.25in]{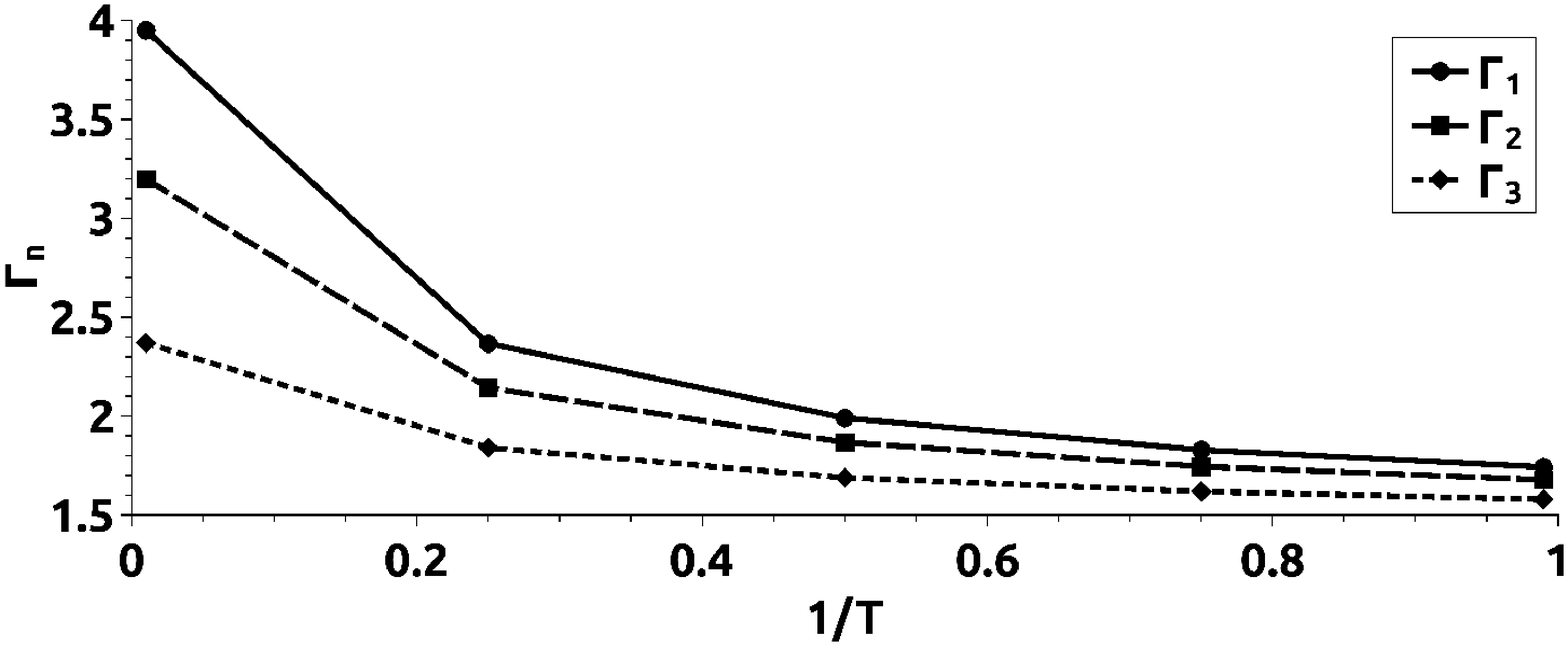}
  \caption{$\Gamma_n$ with different lengths of messages for $F_s=23$ and $F_p=16$}\label{Gn1}
\end{figure}

In our simulation, we assumed the time-out period to be equal to that corresponding to twice the number of attempts for allocating the required number of channels as predicted by our theoretical estimate. With this time-out period, our algorithm always terminated successfully. Accordingly, we set the value of $\sigma$, which is the probability per unit time that the Algorithm~\ref{algo3} terminates unsuccessfully after the time-out period, as equal to zero. For an extremely heavy traffic load of above $96\%$ with $700$ nodes using $FDM-FDMA$ technique, we use the data taken from Table~\ref{t2} with $F_p=23$ and $F_s=16$. Assuming $\sigma = 0$, $\lambda = 0.3$ and $\mu_1 = \mu_2 = \mu_3 = 0.7$. The values of both $P_n$ and $\Gamma_n$ are shown in Tables~\ref{t4}, for $n= 1, ~2, ~3$ and for different values of message length. The values of $P_n$ and $\Gamma_n$ are also shown graphically in Figs.~\ref{Pn1} and~\ref{Gn1}, respectively. From Table~\ref{t4}, we observe that the probability for the system to be in active condition is more for larger lengths of messages. The value of $P_n$ actually depends on two main factors, $i)$ length of the message, and $ii)$ channel mobility (which depends on deallocation of a channel when asked by a primary user and then how quickly we get another channel for communication). Thus, under extremely heavy traffic load of above $95\%$, the probability that the system is in the active condition is effectively dependent only on the length of the message. When the message length is very long, our proposed protocol enables the system to be more than $96\%$ of the time in active condition, while for very short length messages, the system is in active condition for more than $36\%$ of the time. This may be explained by the fact that longer time is needed for transmitting longer messages, and thus the system remains in active condition by grabbing the channel for a larger fraction of time for longer messages.

\begin{table}[]
\caption{{\small $P_n$ and $\Gamma_n$ with different message lengths, $F_s=11$ and $F_p=10$ for overlapped channels with $1000$ nodes}}\label{t41}
\centering
\scriptsize
\begin{tabular}{|c|c|l|l|l|l|l|}
\hline
\multicolumn{2}{|c|}{\multirow{2}{*}{$P_n / \Gamma_n$}} & \multicolumn{5}{|c|}{{$1 / T$}}  \\ \cline{3-7}

\multicolumn{2}{|c|}{} & $0.01$ & $0.25$ & $0.5$ & $0.75$ & $0.99$ \\ \hline
\hline

\multirow{3}{*}{\parbox{1.25cm}{Active State}} & $P_1$ & $0.9742$             & $0.6746$             & $0.5349$          & $0.4464$           & $0.386$ \\ \cline{2-7}
                                                & $P_2$ & $0.9643$             & $0.6374$             & $0.508$           & $0.4266$           & $0.3709$ \\ \cline{2-7}
                                                & $P_3$ & $0.9555$             & $0.6139$             & $0.4922$          & $0.4155$           & $0.3625$ \\ \hline
\hline
 \multirow{3}{*}{\parbox{1.5cm}{Average Waiting Time (in units of $T$)}} & $\Gamma_1$ & $2.6496$ & $1.9298$ & $1.7391$ & $1.6535$ & $1.6065$ \\ \cline{2-7}
                                                                          & $\Gamma_2$ & $3.7059$ & $2.2755$ & $1.9374$ & $1.7918$ & $1.7135$ \\ \cline{2-7}
                                                                          & $\Gamma_3$ & $4.6541$ & $2.5161$ & $2.0633$ & $1.8757$ & $1.7766$ \\
\hline
\end{tabular}
\end{table}

\begin{figure}[]
  \centering
  \includegraphics[width=3.0in,height=2.25in]{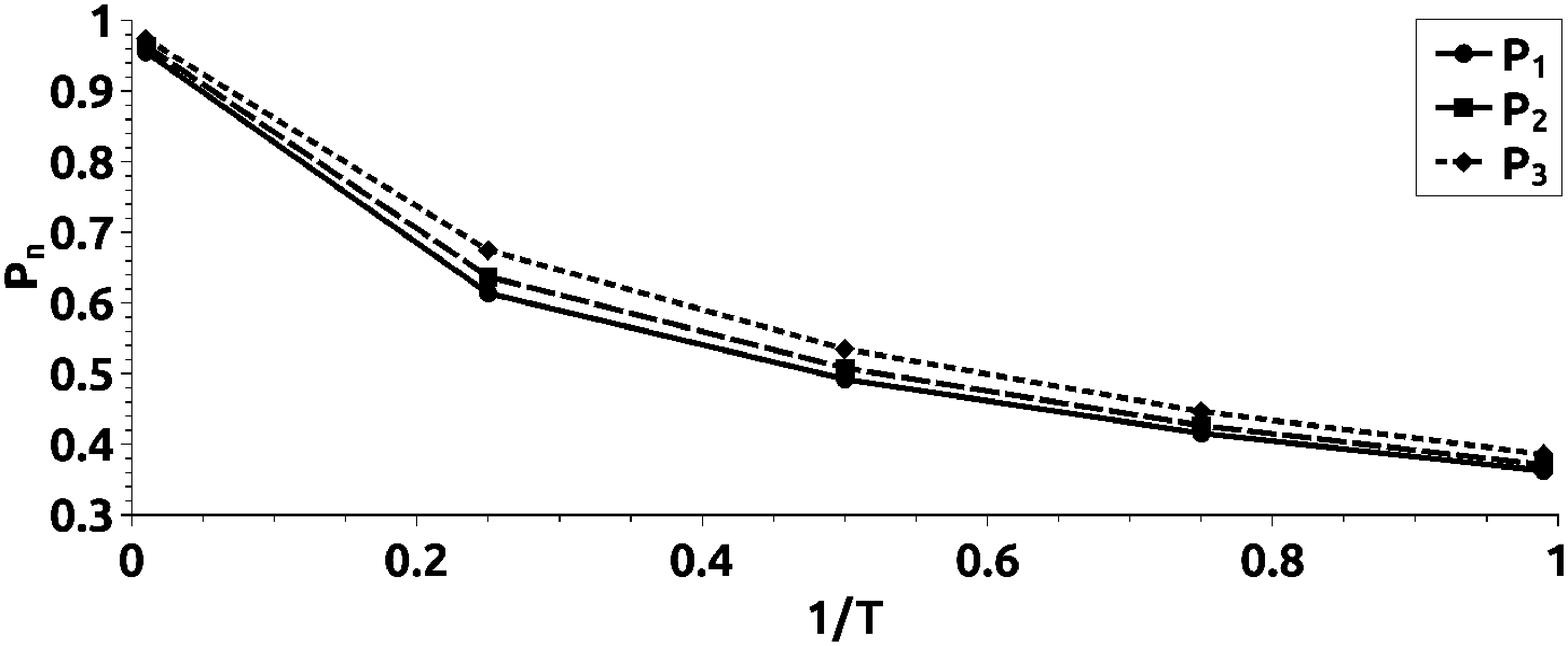}
  \caption{$P_n$ with different lengths of messages for $F_s=11$ and $F_p=10$}\label{Pn2}
\end{figure}

\begin{figure}[]
  \centering
  \includegraphics[width=3.0in,height=2.25in]{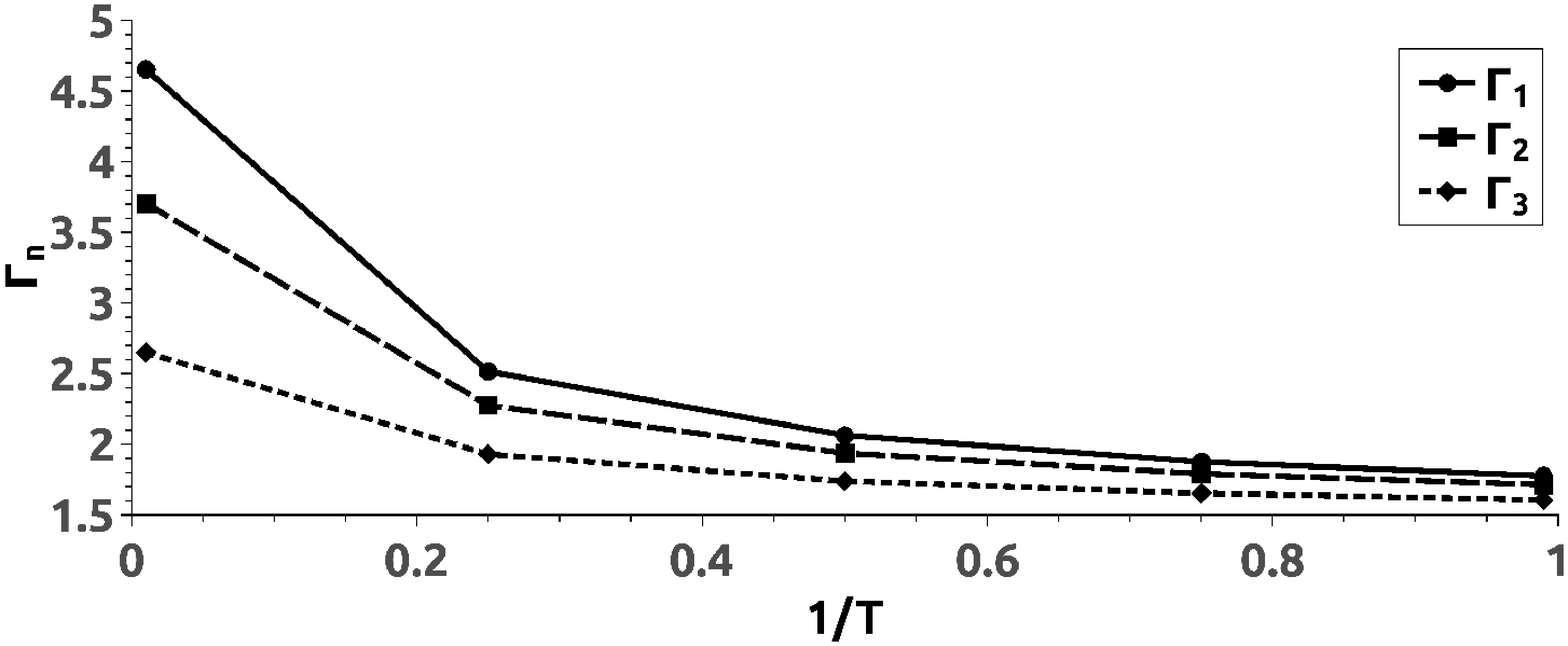}
  \caption{$\Gamma_n$ with different lengths of messages for $F_s=11$ and $F_p=10$}\label{Gn2}
\end{figure}

With $1000$ nodes using  $OFDM-FDMA$ technique for $F_s=11$ and $F_p=10$, we have similarly computed the values of $P_n$ and $\Gamma_n$ for $n = 1, ~2, ~3$ which are shown in Tables~\ref{t41}. These values of $P_n$ and $\Gamma_n$ are also plotted in Figs.~\ref{Pn2} and~\ref{Gn2}, respectively.

\subsection{Generalization to n-Channel System}

To get the value of $P_n$ for any value of $n$, we need five basic types of states for which the probabilities are given as below.
\begin{enumerate}

\item When all reserved channels are secondary :
\begin{equation}
\frac{1}{T} p_n^0 = \sum_{k=1}^{n} \frac{{F_{s} \choose k}}{{F \choose k}} \mu_k p_{n-k}^0
\end{equation}

\item When all reserved channels are primary :
\begin{equation}
\left(\lambda + \frac{1}{T} \right) p_0^n = \sum_{k=1}^{n} \frac{{F_{p} \choose k}}{{F \choose k}} \mu_k p_0^{n-k}
\end{equation}

\item When $k+k'=n$ and $ \forall i,~ 0 < i < n $,
\begin{align}
\left(\lambda + \frac{1}{T} \right) p_{i}^{n-i} &=
  {\sum_{k=0}^{i} \sum_{k'=0 \atop {~k+k' < n}}^{n-i}}  \nonumber \\ &\qquad {}  \left \{ \frac{{F_{p} \choose n-i-k'}{F_{s}
\choose i-k}}{{F \choose n-k-k'}} \mu_{n-k-k'} p_{k}^{k'} \right \}
\end{align}


\item When $k+k'<n$ and $k'=0$ (no primary channel is reserved),
\begin{equation}
\left\{ \sigma + \sum\limits_{i=0}^{n-k} \frac{{F_{p} \choose i} {F_{s} \choose n-k-i}}{{F \choose n-k}} \mu_{n-k} \right\}p_k^0 = \lambda p_k^1
\end{equation}
and

\item When $k+k'<n$ and $k'>0$ (at least one primary channel is reserved),
\begin{align}
\left\{ \lambda + \sigma + \sum\limits_{i=0}^{n-k-k'} \frac{{F_{p} \choose i}{F_{s} \choose n-k-k'-i}}{{F \choose n-k-k'}} \mu_{n-k-k'} \right\} p_k^{k'} &= \nonumber\\ \lambda p_k^{k'+1}
\end{align}


\end{enumerate}

\noindent
Since the system must be always in one of the above types of states, we have,
\begin{equation}
\sum_{k=0}^{n} \sum_{k'=0}^{n-k} p_{k}^{k'} = 1
\end{equation}
When $k+k'=n$, the system will be in {\em active} state, and the probability for the system being in such a state is given by, $P_n = \sum_{k=0}^{k'=n-k} p_{k}^{n-k}$. The average waiting time for an $n-channel$ system can be expressed as $\Gamma_n = T(\frac{1-P_n}{P_n})$.

\section{Conclusion}
A novel channel allocation technique for multimedia communication in a $CRN$ has been presented in this paper. The proposed technique works even when the white spaces in the spectrum do not provide a contiguous bandwidth large enough for maintaining the $QoS$ of the multimedia signals. Our technique is based on first finding a set of non-contiguous white spaces whose total width will be equal to the required bandwidth of the multimedia signal. We then sub-divide the bits from the original signal in the time domain, form sub-packets with these subsets of bits and transmit these sub-packets through the set of channels so found. The algorithms for sensing, allocating and deallocating the required channels from the available white spaces taking into account the presence of high-priority primary users have been presented along with the algorithms for transmitting and receiving the data packets with two different implementations using $FDM-FDMA$ and $OFDM-FDMA$ techniques. The performance of this system has been analyzed by means of a Markov model. Also we find that the average number of attempts for acquiring the required number of channels as obtained from simulation agrees well to the theoretical values for all types of traffic situations ranging from light to extremely heavy (about $96\%$ blocked channels). Simulation results show that the proposed technique always outperforms the existing first-fit and best-fit allocation techniques in terms of the average number of attempts needed for acquiring the necessary number of channels for all traffic situations ranging from light to extremely heavy traffic.

\begin{landscape}
\begin{table}[]
\caption{{\small Average number of attempts required for allocating channels for different types of multimedia signal with $500$ and $700$ nodes in $FDM-FDMA$}}\label{t3}
\centering
\scriptsize
\begin{tabular}{ |c|c|c|c|c|c|c|c|c|c|c|c|c|c|c| }
 \hline
\multirow{2}{*}{\parbox{1.0cm}{$~$ \\Average number of free channels ($F$)\\}} &
\multirow{2}{*}{\parbox{1.0cm}{$~$ \\Demand number of channels ($DN$) \\}} &
\multicolumn{5}{c|}{{\bf Our proposed protocol}} & \multicolumn{4}{c|}{{\bf first-fit protocol}} & \multicolumn{4}{c|}{{\bf best-fit protocol}}\\ \cline{3-15}
& & \parbox{1.0cm}{$~$ \\Number of attempts (theoretical value) \\} &
\parbox{1.0cm}{$~$ \\Number of attempts (simulation results) \\} &
\parbox{1.0cm}{$~$ \\Number of selected primary channels\\} &
\parbox{1.0cm}{$~$ \\Number of selected secondary channels\\} &
\parbox{1.0cm}{$~$ \\Average success rate} &
\parbox{1.0cm}{$~$ \\Number of attempts (simulation results) \\} &
\parbox{1.0cm}{$~$ \\Number of selected primary channels\\} &
\parbox{1.0cm}{$~$ \\Number of selected secondary channels\\} &
\parbox{1.0cm}{$~$ \\Average success rate} &
\parbox{1.0cm}{$~$ \\Number of attempts (simulation results) \\} &
\parbox{1.0cm}{$~$ \\Number of selected primary channels\\} &
\parbox{1.0cm}{$~$ \\Number of selected secondary channels\\} &
\parbox{1.0cm}{$~$ \\Average success rate}\\
 \hline
 \hline

 \multicolumn{15}{|c|}{{\bf For $500$ nodes}}\\
 \hline
 \hline

\multirow{5}{*}{$746$} & $8$ & $2$ & $2$ & $3$ & $5$ & $100$ & $36$ & $4$ & $4$ & $100$ & $160$ & $4$ & $4$ & $100$ \\
                       & $6$ & $2$ & $2$ & $2$ & $4$ & $100$ & $18$ & $3$ & $3$ & $100$ & $88$  & $3$ & $3$ & $100$ \\
                       & $4$ & $2$ & $2$ & $2$ & $2$ & $100$ & $8$  & $2$ & $2$ & $100$ & $48$  & $2$ & $2$ & $100$ \\
                       & $2$ & $2$ & $2$ & $1$ & $1$ & $100$ & $3$  & $1$ & $1$ & $100$ & $26$  & $1$ & $1$ & $100$ \\
                       & $1$ & $2$ & $2$ & $0$ & $1$ & $100$ & $1$  & $0$ & $1$ & $100$ & $19$  & $0$ & $1$ & $100$ \\
 \hline

\multirow{5}{*}{$613$} & $8$ & $2$ & $3$ & $3$ & $5$ & $100$ & $124$ & $4$ & $4$ & $100$ & $315$ & $4$ & $4$ & $100$ \\
                       & $6$ & $2$ & $3$ & $2$ & $4$ & $100$ & $45$  & $3$ & $3$ & $100$ & $124$ & $3$ & $3$ & $100$ \\
                       & $4$ & $2$ & $3$ & $2$ & $2$ & $100$ & $15$  & $2$ & $2$ & $100$ & $45$  & $2$ & $2$ & $100$ \\
                       & $2$ & $2$ & $2$ & $1$ & $1$ & $100$ & $4$   & $1$ & $1$ & $100$ & $16$  & $1$ & $1$ & $100$ \\
                       & $1$ & $2$ & $2$ & $0$ & $1$ & $100$ & $1$   & $0$ & $1$ & $100$ & $9$   & $0$ & $1$ & $100$ \\
 \hline

\multirow{5}{*}{$453$} & $8$ & $3$ & $3$ & $3$ & $5$ & $100$ & $430$ & $4$ & $4$ & $62.266$ & $648$ & $4$ & $4$ & $62.266$ \\
                       & $6$ & $3$ & $3$ & $2$ & $4$ & $100$ & $199$ & $3$ & $3$ & $99.498$ & $350$ & $3$ & $3$ & $99.498$ \\
                       & $4$ & $3$ & $3$ & $2$ & $2$ & $100$ & $40$  & $2$ & $2$ & $100$    & $77$  & $2$ & $2$ & $100$    \\
                       & $2$ & $3$ & $3$ & $1$ & $1$ & $100$ & $7$   & $1$ & $1$ & $100$    & $15$  & $1$ & $1$ & $100$    \\
                       & $1$ & $3$ & $3$ & $0$ & $1$ & $100$ & $2$   & $0$ & $1$ & $100$    & $6$   & $0$ & $1$ & $100$    \\
 \hline

\multirow{5}{*}{$279$} & $8$ & $4$ & $5$ & $3$ & $5$ & $100$ & $508$ & $4$ & $4$ & $2.441$  & $630$ & $4$ & $4$ & $2.441$  \\
                       & $6$ & $4$ & $5$ & $2$ & $4$ & $100$ & $476$ & $3$ & $3$ & $28.429$ & $610$ & $3$ & $3$ & $28.429$ \\
                       & $4$ & $4$ & $4$ & $1$ & $3$ & $100$ & $213$ & $2$ & $2$ & $99.136$ & $292$ & $2$ & $2$ & $99.136$ \\
                       & $2$ & $4$ & $4$ & $1$ & $1$ & $100$ & $16$  & $1$ & $1$ & $100$    & $23$  & $1$ & $1$ & $100$    \\
                       & $1$ & $4$ & $4$ & $0$ & $1$ & $100$ & $3$   & $0$ & $1$ & $100$    & $5$   & $0$ & $1$ & $100$    \\
 \hline
 \hline

 \multicolumn{15}{|c|}{{\bf For $700$ nodes}}\\

 \hline
 \hline

\multirow{5}{*}{$697$} & $8$ & $2$ & $2$ & $3$ & $5$ & $100$ & $55$ & $4$ & $4$ & $100$ & $192$ & $3$ & $5$ & $100$ \\
                       & $6$ & $2$ & $2$ & $2$ & $4$ & $100$ & $25$ & $3$ & $3$ & $100$ & $93$  & $2$ & $4$ & $100$ \\
                       & $4$ & $2$ & $2$ & $2$ & $2$ & $100$ & $10$ & $2$ & $2$ & $100$ & $44$  & $2$ & $2$ & $100$ \\
                       & $2$ & $2$ & $2$ & $1$ & $1$ & $100$ & $3$  & $1$ & $1$ & $100$ & $21$  & $1$ & $1$ & $100$ \\
                       & $1$ & $2$ & $2$ & $0$ & $1$ & $100$ & $1$  & $0$ & $1$ & $100$ & $14$  & $0$ & $1$ & $100$ \\
 \hline

\multirow{5}{*}{$510$} & $8$ & $2$ & $3$ & $3$ & $5$ & $100$ & $332$ & $4$ & $4$ & $90.594$ & $570$ & $3$ & $5$ & $90.594$ \\
                       & $6$ & $2$ & $3$ & $2$ & $4$ & $100$ & $112$ & $3$ & $3$ & $99.996$ & $230$ & $2$ & $4$ & $99.996$ \\
                       & $4$ & $2$ & $3$ & $2$ & $2$ & $100$ & $27$  & $2$ & $2$ & $100$    & $60$  & $2$ & $2$ & $100$    \\
                       & $2$ & $2$ & $3$ & $1$ & $1$ & $100$ & $5$   & $1$ & $1$ & $100$    & $14$  & $1$ & $1$ & $100$    \\
                       & $1$ & $2$ & $2$ & $0$ & $1$ & $100$ & $1$   & $0$ & $1$ & $100$    & $7$   & $0$ & $1$ & $100$    \\
 \hline

\multirow{5}{*}{$285$} & $8$ & $4$ & $4$ & $3$ & $5$ & $100$ & $500$ & $4$ & $4$ & $2.844$  & $638$ & $3$ & $5$ & $2.844$  \\
                       & $6$ & $4$ & $4$ & $2$ & $4$ & $100$ & $474$ & $3$ & $3$ & $31.158$ & $610$ & $3$ & $3$ & $31.158$ \\
                       & $4$ & $4$ & $4$ & $2$ & $2$ & $100$ & $200$ & $2$ & $2$ & $99.468$ & $277$ & $1$ & $3$ & $99.468$ \\
                       & $2$ & $4$ & $4$ & $1$ & $1$ & $100$ & $15$  & $1$ & $1$ & $100$    & $22$  & $1$ & $1$ & $100$    \\
                       & $1$ & $4$ & $4$ & $0$ & $1$ & $100$ & $3$   & $0$ & $1$ & $100$    & $5$   & $0$ & $1$ & $100$    \\
 \hline

\multirow{5}{*}{$39$} & $8$ & $26$ & $29$ & $3$ & $5$ & $100$ & $0$   & $0$ & $0$ & $0$      & $0$   & $0$ & $0$ & $0$      \\
                      & $6$ & $26$ & $28$ & $3$ & $3$ & $100$ & $0$   & $0$ & $0$ & $0$      & $0$   & $0$ & $0$ & $0$      \\
                      & $4$ & $26$ & $28$ & $2$ & $2$ & $100$ & $494$ & $2$ & $2$ & $0.204$  & $502$ & $2$ & $2$ & $0.204$  \\
                      & $2$ & $26$ & $27$ & $1$ & $1$ & $100$ & $391$ & $1$ & $1$ & $78.472$ & $406$ & $1$ & $2$ & $78.472$ \\
                      & $1$ & $26$ & $26$ & $0$ & $1$ & $100$ & $25$  & $0$ & $1$ & $100$    & $26$  & $0$ & $1$ & $100$    \\
 \hline

\end{tabular}
\end{table}
\end{landscape}

\begin{landscape}
\begin{table}[]
\caption{{\small Average number of attempts required for allocating channels for different types of multimedia signal with $700$ and $1000$ nodes in $OFDM-FDMA$}}\label{t3_1}
\centering
\scriptsize
\begin{tabular}{ |c|c|c|c|c|c|c|c|c|c|c|c|c|c|c| }
 \hline
\multirow{2}{*}{\parbox{1.0cm}{$~$ \\Average number of free channels ($F$)\\}} &
\multirow{2}{*}{\parbox{1.0cm}{$~$ \\Demand number of channels ($DN$) \\}} &
\multicolumn{5}{c|}{{\bf Our proposed protocol}} & \multicolumn{4}{c|}{{\bf first-fit protocol}} & \multicolumn{4}{c|}{{\bf best-fit protocol}}\\ \cline{3-15}
& & \parbox{1.0cm}{$~$ \\Number of attempts (theoretical value) \\} &
\parbox{1.0cm}{$~$ \\Number of attempts$^*$ (simulation results) \\} &
\parbox{1.0cm}{$~$ \\Number of selected primary channels\\} &
\parbox{1.0cm}{$~$ \\Number of selected secondary channels\\} &
\parbox{1.0cm}{$~$ \\Average success rate} &
\parbox{1.0cm}{$~$ \\Number of attempts (simulation results) \\} &
\parbox{1.0cm}{$~$ \\Number of selected primary channels\\} &
\parbox{1.0cm}{$~$ \\Number of selected secondary channels\\} &
\parbox{1.0cm}{$~$ \\Average success rate} &
\parbox{1.0cm}{$~$ \\Number of attempts (simulation results) \\} &
\parbox{1.0cm}{$~$ \\Number of selected primary channels\\} &
\parbox{1.0cm}{$~$ \\Number of selected secondary channels\\} &
\parbox{1.0cm}{$~$ \\Average success rate}\\
 \hline
 \hline

 \multicolumn{15}{|c|}{{\bf For $700$ nodes}}\\
 \hline
 \hline

\multirow{5}{*}{$1185$} & $8$ & $1$ & $2$ & $4$ & $4$ & $100$ & $42$ & $4$ & $4$ & $100$ & $229$ & $4$ & $4$ & $100$ \\
                        & $6$ & $1$ & $2$ & $3$ & $3$ & $100$ & $30$ & $3$ & $3$ & $100$ & $161$ & $3$ & $3$ & $100$ \\
                        & $4$ & $1$ & $2$ & $2$ & $2$ & $100$ & $16$ & $2$ & $2$ & $100$ & $109$ & $2$ & $2$ & $100$ \\
                        & $2$ & $1$ & $2$ & $1$ & $1$ & $100$ & $9$  & $1$ & $1$ & $100$ & $74$  & $1$ & $1$ & $100$ \\
                        & $1$ & $1$ & $2$ & $0$ & $1$ & $100$ & $4$  & $0$ & $1$ & $100$ & $63$  & $0$ & $1$ & $100$ \\
 \hline

\multirow{5}{*}{$798$} & $8$ & $1$ & $2$ & $4$ & $4$ & $100$ & $94$ & $4$ & $4$ & $100$ & $323$ & $4$ & $4$ & $100$ \\
                       & $6$ & $1$ & $2$ & $3$ & $3$ & $100$ & $52$ & $3$ & $3$ & $100$ & $174$ & $3$ & $3$ & $100$ \\
                       & $4$ & $1$ & $2$ & $2$ & $2$ & $100$ & $25$ & $2$ & $2$ & $100$ & $90$  & $2$ & $2$ & $100$ \\
                       & $2$ & $1$ & $2$ & $1$ & $1$ & $100$ & $12$ & $1$ & $1$ & $100$ & $47$  & $1$ & $1$ & $100$ \\
                       & $1$ & $1$ & $2$ & $0$ & $1$ & $100$ & $5$  & $0$ & $1$ & $100$ & $31$  & $0$ & $1$ & $100$ \\
 \hline

\multirow{5}{*}{$456$} & $8$ & $2$ & $2$ & $4$ & $4$ & $100$ & $405$ & $4$ & $4$ & $99.473$ & $874$ & $4$ & $4$ & $99.473$ \\
                       & $6$ & $2$ & $2$ & $3$ & $3$ & $100$ & $152$ & $3$ & $3$ & $100$    & $364$ & $3$ & $3$ & $100$    \\
                       & $4$ & $2$ & $2$ & $2$ & $2$ & $100$ & $53$  & $2$ & $2$ & $100$    & $129$ & $2$ & $2$ & $100$    \\
                       & $2$ & $2$ & $2$ & $1$ & $1$ & $100$ & $18$  & $1$ & $1$ & $100$    & $45$  & $1$ & $1$ & $100$    \\
                       & $1$ & $2$ & $2$ & $0$ & $1$ & $100$ & $8$   & $0$ & $1$ & $100$    & $24$  & $0$ & $1$ & $100$    \\
 \hline

\multirow{5}{*}{$211$} & $8$ & $4$ & $4$ & $4$ & $4$ & $100$ & $946$ & $4$ & $4$ & $34.442$ & $1386$ & $4$ & $4$ & $34.442$ \\
                       & $6$ & $4$ & $4$ & $3$ & $3$ & $100$ & $686$ & $3$ & $3$ & $88.625$ & $1084$ & $3$ & $3$ & $88.625$ \\
                       & $4$ & $4$ & $4$ & $2$ & $2$ & $100$ & $191$ & $2$ & $2$ & $100$    & $343$  & $2$ & $2$ & $100$    \\
                       & $2$ & $4$ & $4$ & $1$ & $1$ & $100$ & $39$  & $1$ & $1$ & $100$    & $70$   & $1$ & $1$ & $100$    \\
                       & $1$ & $4$ & $4$ & $0$ & $0$ & $100$ & $15$  & $0$ & $1$ & $100$    & $29$   & $0$ & $1$ & $100$    \\
 \hline
 \hline

 \multicolumn{15}{|c|}{{\bf For $1000$ nodes}}\\

 \hline
 \hline

\multirow{5}{*}{$1015$} & $8$ & $1$ & $2$ & $4$ & $4$ & $100$ & $56$ & $4$ & $4$ & $100$ & $242$ & $4$ & $4$ & $100$ \\
                        & $6$ & $1$ & $2$ & $3$ & $3$ & $100$ & $36$ & $3$ & $3$ & $100$ & $154$ & $3$ & $3$ & $100$ \\
                        & $4$ & $1$ & $2$ & $2$ & $2$ & $100$ & $19$ & $2$ & $2$ & $100$ & $93$  & $2$ & $2$ & $100$ \\
                        & $2$ & $1$ & $2$ & $1$ & $1$ & $100$ & $10$ & $1$ & $1$ & $100$ & $57$  & $1$ & $1$ & $100$ \\
                        & $1$ & $1$ & $2$ & $0$ & $0$ & $100$ & $4$  & $0$ & $1$ & $100$ & $43$  & $0$ & $1$ & $100$ \\
 \hline

\multirow{5}{*}{$546$} & $8$ & $2$ & $2$ & $4$ & $4$ & $100$ & $250$ & $4$ & $4$ & $99.996$ & $639$ & $4$ & $4$ & $99.996$ \\
                       & $6$ & $2$ & $2$ & $3$ & $3$ & $100$ & $104$ & $3$ & $3$ & $100$    & $274$ & $3$ & $3$ & $100$    \\
                       & $4$ & $2$ & $2$ & $2$ & $2$ & $100$ & $41$  & $2$ & $2$ & $100$    & $109$ & $2$ & $2$ & $100$    \\
                       & $2$ & $2$ & $2$ & $1$ & $1$ & $100$ & $16$  & $1$ & $1$ & $100$    & $44$  & $1$ & $1$ & $100$    \\
                       & $1$ & $2$ & $2$ & $0$ & $1$ & $100$ & $7$   & $0$ & $1$ & $100$    & $25$  & $0$ & $1$ & $100$    \\
 \hline

\multirow{5}{*}{$205$} & $8$ & $4$ & $4$ & $4$ & $4$ & $100$ & $958$ & $4$ & $4$ & $31.828$ & $1388$ & $4$ & $4$ & $31.828$\\
                       & $6$ & $4$ & $4$ & $3$ & $3$ & $100$ & $707$ & $3$ & $3$ & $86.829$ & $1103$ & $3$ & $3$ & $86.829$\\
                       & $4$ & $4$ & $4$ & $2$ & $2$ & $100$ & $201$ & $2$ & $2$ & $99.998$ & $358$  & $2$ & $2$ & $99.998$\\
                       & $2$ & $4$ & $4$ & $1$ & $1$ & $100$ & $40$  & $1$ & $1$ & $100$    & $71$   & $1$ & $1$ & $100$   \\
                       & $1$ & $4$ & $4$ & $0$ & $1$ & $100$ & $15$  & $0$ & $1$ & $100$    & $29$   & $0$ & $1$ & $100$   \\
 \hline

\multirow{5}{*}{$21$} & $8$ & $32$ & $45$ & $4$ & $4$ & $100$ & $0$   & $0$ & $0$ & $0$    & $0$    & $0$ & $0$ & $0$    \\
                      & $6$ & $32$ & $41$ & $3$ & $3$ & $100$ & $413$ & $3$ & $3$ & $0.1$  & $413$  & $3$ & $3$ & $0.1$  \\
                      & $4$ & $32$ & $39$ & $2$ & $2$ & $100$ & $965$ & $2$ & $2$ & $5.7$  & $1218$ & $2$ & $2$ & $5.7$  \\
                      & $2$ & $32$ & $37$ & $1$ & $1$ & $100$ & $582$ & $1$ & $1$ & $90.3$ & $658$  & $1$ & $1$ & $90.3$ \\
                      & $1$ & $32$ & $36$ & $0$ & $1$ & $100$ & $115$ & $0$ & $1$ & $100$  & $135$  & $0$ & $1$ & $100$  \\
 \hline

 \multicolumn{15}{l}{$^*$The sensing time per attempts in $OFDM-FDMA$ channel allocation technique is three times more than that in $FDM-FDMA$ channel allocation technique.}

\end{tabular}
\end{table}
\end{landscape}

\begin{landscape}
\begin{table}[]
\caption{Average Delay and Delay Jitter for different real-life Video files} \label{different_video_file}
\scriptsize
\centering
\begin{tabular}{|c|c|c|c|c|c|c|c|c|}
\hline
\parbox{1.75cm}{Video File Size (in bits)} & \parbox{1.75cm}{Average Number of Channel Deallocation by PUs} & \parbox{1.75cm}{Ideal Transmission Time (IT) (in sec)} & \parbox{1.75cm}{Initial Channel Allocation Time (ICA) (in msec)} & \parbox{1.75cm}{Channel Reallocation Time (CR) (in msec)} & \parbox{2cm}{Actual Transmission Time (AT) (in sec) (AT=IT+ICA+CR)} & \parbox{1.75cm}{Maximum Jitter (in msec)} & \parbox{1.75cm}{Mean Jitter (in msec)} & \parbox{1.75cm}{Standard Deviation In Jitter (in msec)} \\ \hline

\multicolumn{9}{|c|}{Average number of free channels ($F$) is $697$ }    \\ \hline
$387685216$ & $3.901$ & $757.1976875$ & $19.8$   & $5.85375$  & $757.22334125$ & $1.96625$  & $0.000989475$  & $0.038375$ \\
$389171680$ & $4.053$ & $760.1009375$ & $19.68$  & $6.0525$   & $760.12667$    & $1.94$     & $0.0010191125$ & $0.038625$ \\
$390823936$ & $4.059$ & $763.328$     & $19.79$  & $6.09375$  & $763.35388375$ & $1.99875$  & $0.00102175$   & $0.039125$ \\
$393725152$ & $3.891$ & $768.9944375$ & $19.92$  & $5.88375$  & $769.02024125$ & $1.97625$  & $0.000979325$  & $0.03825$  \\
$399652192$ & $3.973$ & $780.5706875$ & $19.76$  & $5.9975$   & $780.596445$   & $2.01$     & $0.0009833625$ & $0.0385$   \\
$403520768$ & $4.072$ & $788.1265$    & $19.61$  & $6.22125$  & $788.15233125$ & $2.07625$  & $0.001010275$  & $0.039625$ \\
$411901408$ & $4.083$ & $804.4949375$ & $19.61$  & $6.155$    & $804.5207025$  & $2.01125$  & $0.0009791625$ & $0.038375$ \\
$413877088$ & $4.049$ & $808.3536875$ & $19.76$  & $6.07375$  & $808.37952125$ & $1.97875$  & $0.00096165$   & $0.03775$  \\
$415495232$ & $4.168$ & $811.514125$  & $19.79$  & $6.27875$  & $811.54019375$ & $2.03125$  & $0.0009903375$ & $0.038625$ \\
$420989536$ & $4.112$ & $822.2451875$ & $19.71$  & $6.1375$   & $822.271035$   & $1.99375$  & $0.0009554$    & $0.03775$  \\ 
\boldmath{$402684220.8$} & \boldmath{$4.0361$} & \boldmath{$786.49261875$} & \boldmath{$19.743$} & \boldmath{$6.07475$} & \boldmath{$786.5184365$} & \boldmath{$1.99825$} & \boldmath{$0.000988985$} & \boldmath{$0.0385$} \\ \hline

\multicolumn{9}{|c|}{Average number of free channels ($F$) is $510$ }    \\ \hline
$387685216$ & $4.026$ & $757.1976875$ & $23.68$  & $6.48625$  & $757.22785375$ & $2.31625$  & $0.0010963875$ & $0.043125$ \\
$389171680$ & $3.923$ & $760.1009375$ & $24.26$  & $6.3325$   & $760.13153$    & $2.285$    & $0.0010662625$ & $0.042375$ \\
$390823936$ & $4.011$ & $763.328$     & $24.02$  & $6.4825$   & $763.3585025$  & $2.305$    & $0.0010869375$ & $0.042875$ \\
$393725152$ & $3.986$ & $768.9944375$ & $24.14$  & $6.51125$  & $769.02508875$ & $2.35625$  & $0.0010837625$ & $0.043125$ \\
$399652192$ & $4.191$ & $780.5706875$ & $23.93$  & $6.7525$   & $780.60137$    & $2.34625$  & $0.0011071$    & $0.0435$   \\
$403520768$ & $4.13$  & $788.1265$    & $24.24$  & $6.64625$  & $788.15738625$ & $2.3125$   & $0.0010792875$ & $0.0425$   \\
$411901408$ & $4.147$ & $804.4949375$ & $23.96$  & $6.675$    & $804.5255725$  & $2.32$     & $0.00106205$   & $0.042375$ \\
$413877088$ & $4.071$ & $808.3536875$ & $24.18$  & $6.5425$   & $808.38441$    & $2.2825$   & $0.0010358625$ & $0.0415$   \\
$415495232$ & $4.278$ & $811.514125$  & $24$     & $6.80875$  & $811.54493375$ & $2.3025$   & $0.0010739375$ & $0.042375$ \\
$420989536$ & $4.139$ & $822.2451875$ & $23.8$   & $6.80375$  & $822.27579125$ & $2.4$      & $0.0010591125$ & $0.042875$ \\
\boldmath{$402684220.8$} & \boldmath{$4.0902$} & \boldmath{$786.49261875$} & \boldmath{$24.021$} & \boldmath{$6.604125$} & \boldmath{$786.523243875$} & \boldmath{$2.322625$} & \boldmath{$0.001075075$} & \boldmath{$0.0426625$} \\ \hline

\multicolumn{9}{|c|}{Average number of free channels ($F$) is $285$ }    \\ \hline
$387685216$ & $4.293$ & $757.1976875$ & $39.77$  & $9.4225$   & $757.24688$    & $3.9$      & $0.001625$ & $0.06575$  \\
$389171680$ & $4.202$ & $760.1009375$ & $39.83$  & $9.1$      & $760.1498675$  & $3.8725$   & $0.0015$   & $0.064125$ \\
$390823936$ & $4.305$ & $763.328$     & $40.04$  & $9.185$    & $763.377225$   & $3.8$      & $0.0015$   & $0.0635$   \\
$393725152$ & $4.306$ & $768.9944375$ & $39.41$  & $9.32625$  & $769.04317375$ & $3.92875$  & $0.0015$   & $0.064875$ \\
$399652192$ & $4.393$ & $780.5706875$ & $40.35$  & $9.3875$   & $780.620425$   & $3.94875$  & $0.0015$   & $0.064375$ \\
$403520768$ & $4.212$ & $788.1265$    & $39.82$  & $9.16875$  & $788.17548875$ & $3.9225$   & $0.0015$   & $0.063875$ \\
$411901408$ & $4.409$ & $804.4949375$ & $39.88$  & $9.3675$   & $804.544185$   & $3.81375$  & $0.0015$   & $0.0625$   \\
$413877088$ & $4.25$  & $808.3536875$ & $39.76$  & $9.2375$   & $808.402685$   & $3.8975$   & $0.0015$   & $0.06275$  \\
$415495232$ & $4.323$ & $811.514125$  & $40.55$  & $9.25375$  & $811.56392875$ & $3.88$     & $0.0015$   & $0.0625$   \\
$420989536$ & $4.394$ & $822.2451875$ & $40.11$  & $9.5325$   & $822.29483$    & $4.01125$  & $0.0015$   & $0.06375$  \\ 
\boldmath{$402684220.8$} & \boldmath{$4.3087$} & \boldmath{$786.49261875$} & \boldmath{$39.952$} & \boldmath{$9.298125$} & \boldmath{$786.541868875$} & \boldmath{$3.8975$} & \boldmath{$0.0015125$} & \boldmath{$0.0638$} \\ \hline

\multicolumn{9}{|c|}{Average number of free channels ($F$) is $39$ }    \\ \hline
$387685216$ & $5.169$ & $757.1976875$ & $279.59$ & $70.92625$ & $757.54820375$ & $38.32625$ & $0.012$    & $0.578125$ \\
$389171680$ & $4.919$ & $760.1009375$ & $281.79$ & $70.46875$ & $760.45319625$ & $38.9775$  & $0.011875$ & $0.581$    \\
$390823936$ & $4.971$ & $763.328$     & $286.82$ & $69.2$     & $763.68402$    & $37.66625$ & $0.011625$ & $0.564125$ \\
$393725152$ & $5.074$ & $768.9944375$ & $283.47$ & $70.0275$  & $769.347935$   & $36.8475$  & $0.011625$ & $0.55675$  \\
$399652192$ & $5.119$ & $780.5706875$ & $288.57$ & $68.0825$  & $780.92734$    & $36.27125$ & $0.011125$ & $0.5405$   \\
$403520768$ & $5.016$ & $788.1265$    & $282.11$ & $66.82875$ & $788.47543875$ & $36.2325$  & $0.010875$ & $0.534625$ \\
$411901408$ & $5.083$ & $804.4949375$ & $285.96$ & $70.90625$ & $804.85180375$ & $37.86625$ & $0.01125$  & $0.55925$  \\
$413877088$ & $5.077$ & $808.3536875$ & $281.94$ & $68.72875$ & $808.70435625$ & $37.84625$ & $0.010875$ & $0.548125$ \\
$415495232$ & $5.093$ & $811.514125$  & $285.01$ & $67.725$   & $811.86686$    & $36.2175$  & $0.010625$ & $0.528875$ \\
$420989536$ & $5.16$  & $822.2451875$ & $281.46$ & $68.43625$ & $822.59508375$ & $36.64125$ & $0.010625$ & $0.531625$ \\ 
\boldmath{$402684220.8$} & \boldmath{$5.0681$} & \boldmath{$786.49261875$} & \boldmath{$283.672$} & \boldmath{$69.133$} & \boldmath{$786.84542375$} & \boldmath{$37.28925$} & \boldmath{$0.01125$} & \boldmath{$0.5523$} \\ \hline

\end{tabular}
\end{table}
\end{landscape}

\begin{landscape}
\begin{table}[]
\caption{Average Delay and Delay Jitter for different real-life Music files} \label{different_music_file}
\scriptsize
\centering
\begin{tabular}{|c|c|c|c|c|c|c|c|c|}
\hline
\parbox{1.75cm}{Music File Size (in bits)} & \parbox{1.75cm}{Average Number of Channel Deallocation by PUs} & \parbox{1.75cm}{Ideal Transmission Time (IT) (in sec)} & \parbox{1.75cm}{Initial Channel Allocation Time (ICA) (in msec)} & \parbox{1.75cm}{Channel Reallocation Time (CR) (in msec)} & \parbox{2cm}{Actual Transmission Time (AT) (in sec) (AT=IT+ICA+CR)} & \parbox{1.75cm}{Maximum Jitter (in msec)} & \parbox{1.75cm}{Mean Jitter (in msec)} & \parbox{1.75cm}{Standard Deviation In Jitter (in msec)} \\ \hline

\multicolumn{9}{|c|}{Average number of free channels ($F$) is $697$ }    \\ \hline
$48070656$ & $0.745$  & $125.184$        & $14.2575$ & $1.14625$ & $125.19940375$   & $0.8475$  & $0.0011720375$ & $0.03025$  \\
$48381952$ & $0.785$  & $125.9946666667$ & $14.475$  & $1.22125$ & $126.0103629167$ & $0.9125$  & $0.00123985$   & $0.03225$  \\
$48382976$ & $0.822$  & $125.9973333333$ & $14.55$   & $1.265$   & $126.0131483333$ & $0.92375$ & $0.00125$      & $0.033$    \\
$48451864$ & $0.8$    & $126.1767291667$ & $14.3175$ & $1.27125$ & $126.1923179167$ & $0.9475$  & $0.00125$      & $0.0335$   \\
$48594944$ & $0.787$  & $126.5493333333$ & $14.4825$ & $1.19125$ & $126.5650070833$ & $0.8625$  & $0.0012045$    & $0.030875$ \\
$48647680$ & $0.824$  & $126.6866666667$ & $14.4375$ & $1.285$   & $126.7023891667$ & $0.93$    & $0.00125$      & $0.03325$  \\
$49990032$ & $0.785$  & $130.182375$     & $14.4525$ & $1.2275$  & $130.198055$     & $0.885$   & $0.0012058$    & $0.031125$ \\
$50430976$ & $0.831$  & $131.3306666667$ & $14.3925$ & $1.32$    & $131.3463791667$ & $0.95875$ & $0.00125$      & $0.0335$   \\
$50432936$ & $0.842$  & $131.3357708333$ & $14.3925$ & $1.33$    & $131.3514933333$ & $0.98375$ & $0.00125$      & $0.034125$ \\
$51168256$ & $0.899$  & $133.2506666667$ & $14.505$  & $1.375$   & $133.2665466667$ & $0.97125$ & $0.001375$     & $0.034125$ \\ 
\boldmath{$49255227.2$} & \boldmath{$0.812$} & \boldmath{$128.2688208333$} & \boldmath{$14.42625$} & \boldmath{$1.26325$} & \boldmath{$128.2845103333$} & \boldmath{$0.92225$} & \boldmath{$0.0012447188$} & \boldmath{$0.0326$} \\ \hline

\multicolumn{9}{|c|}{Average number of free channels ($F$) is $510$ }    \\ \hline
$48070656$ & $0.921$  & $125.184$        & $17.7375$ & $1.76125$ & $125.20349875$   & $1.32375$ & $0.00175$  & $0.0465$  \\
$48381952$ & $0.946$  & $125.9946666667$ & $18.12$   & $1.78875$ & $126.0145754167$ & $1.3025$  & $0.001875$ & $0.04575$ \\
$48382976$ & $0.947$  & $125.9973333333$ & $17.9175$ & $1.87375$ & $126.0171245833$ & $1.3825$  & $0.001875$ & $0.04825$ \\
$48451864$ & $0.997$  & $126.1767291667$ & $17.9625$ & $1.93875$ & $126.1966304167$ & $1.36875$ & $0.002$    & $0.04875$ \\
$48594944$ & $0.919$  & $126.5493333333$ & $17.9925$ & $1.79625$ & $126.5691220833$ & $1.30875$ & $0.001875$ & $0.04575$ \\
$48647680$ & $0.939$  & $126.6866666667$ & $17.7075$ & $1.775$   & $126.7061491667$ & $1.2975$  & $0.00175$  & $0.0455$  \\
$49990032$ & $0.993$  & $130.182375$     & $17.7$    & $1.87375$ & $130.20194875$   & $1.33625$ & $0.001875$ & $0.04675$ \\
$50430976$ & $1.006$  & $131.3306666667$ & $17.67$   & $1.865$   & $131.3502016667$ & $1.37$    & $0.001875$ & $0.04725$ \\
$50432936$ & $0.983$  & $131.3357708333$ & $17.8425$ & $1.8775$  & $131.3554908333$ & $1.31875$ & $0.001875$ & $0.046$   \\
$51168256$ & $0.99$   & $133.2506666667$ & $17.8125$ & $1.83$    & $133.2703091667$ & $1.3225$  & $0.00175$  & $0.0455$  \\ 
\boldmath{$49255227.2$} & \boldmath{$0.9641$} & \boldmath{$128.2688208333$} & \boldmath{$17.84625$} & \boldmath{$1.838$} & \boldmath{$128.2885050833$} & \boldmath{$1.333125$} & \boldmath{$0.00185$} & \boldmath{$0.0466$} \\ \hline

\multicolumn{9}{|c|}{Average number of free channels ($F$) is $285$ }    \\ \hline
$48070656$ & $1.338$  & $125.184$        & $29.5875$  & $4.1925$   & $125.21778$      & $2.92375$  & $0.00425$  & $0.103375$ \\
$48381952$ & $1.344$  & $125.9946666667$ & $29.9175$  & $4.015$    & $126.0285991667$ & $2.7925$   & $0.004125$ & $0.0985$   \\
$48382976$ & $1.365$  & $125.9973333333$ & $29.37$    & $4.1725$   & $126.0308758333$ & $2.92625$  & $0.00425$  & $0.102875$ \\
$48451864$ & $1.353$  & $126.1767291667$ & $29.2425$  & $4.2225$   & $126.2101941667$ & $2.99375$  & $0.00425$  & $0.104625$ \\
$48594944$ & $1.345$  & $126.5493333333$ & $29.16$    & $4.1525$   & $126.5826458333$ & $2.9225$   & $0.00425$  & $0.10225$  \\
$48647680$ & $1.395$  & $126.6866666667$ & $29.7525$  & $4.205$    & $126.7206241667$ & $2.905$    & $0.00425$  & $0.1025$   \\
$49990032$ & $1.429$  & $130.182375$     & $29.1$     & $4.49$     & $130.215965$     & $3.12875$  & $0.004375$ & $0.108125$ \\
$50430976$ & $1.432$  & $131.3306666667$ & $29.4225$  & $4.41375$  & $131.3645029167$ & $3.1$      & $0.00425$  & $0.10625$  \\
$50432936$ & $1.334$  & $131.3357708333$ & $29.79$    & $4.11375$  & $131.3696745833$ & $2.91125$  & $0.004$    & $0.100125$ \\
$51168256$ & $1.373$  & $133.2506666667$ & $29.1$     & $4.27875$  & $133.2840454167$ & $3.0225$   & $0.004125$ & $0.10275$  \\ 
\boldmath{$49255227.2$} & \boldmath{$1.3708$} & \boldmath{$128.2688208333$} & \boldmath{$29.44425$} & \boldmath{$4.225625$} & \boldmath{$128.3024907083$} & \boldmath{$2.962625$} & \boldmath{$0.0042125$} & \boldmath{$0.1031375$} \\ \hline

\multicolumn{9}{|c|}{Average number of free channels ($F$) is $39$ }    \\ \hline
$48070656$ & $2.811$  & $125.184$        & $208.0425$ & $46.91$    & $125.4389525$    & $29.3175$  & $0.048$    & $1.05175$  \\
$48381952$ & $2.821$  & $125.9946666667$ & $205.545$  & $48.30125$ & $126.2485129167$ & $30.1775$  & $0.049$    & $1.07725$  \\
$48382976$ & $2.798$  & $125.9973333333$ & $209.4075$ & $46.585$   & $126.2533258333$ & $29.2625$  & $0.04725$  & $1.042625$ \\
$48451864$ & $2.695$  & $126.1767291667$ & $209.3625$ & $43.43375$ & $126.4295254167$ & $27.34125$ & $0.044$    & $0.974$    \\
$48594944$ & $2.794$  & $126.5493333333$ & $210.39$   & $47.80125$ & $126.8075245833$ & $29.93875$ & $0.048375$ & $1.067125$ \\
$48647680$ & $2.899$  & $126.6866666667$ & $207.0825$ & $46.5275$  & $126.9402766667$ & $29.435$   & $0.047$    & $1.04275$  \\
$49990032$ & $2.821$  & $130.182375$     & $205.575$  & $48.3425$  & $130.4362925$    & $30.51$    & $0.0475$   & $1.068125$ \\
$50430976$ & $2.839$  & $131.3306666667$ & $211.38$   & $46.33125$ & $131.5883779167$ & $29.09875$ & $0.045125$ & $1.01725$  \\
$50432936$ & $2.93$   & $131.3357708333$ & $206.445$  & $47.4175$  & $131.5896333333$ & $29.33375$ & $0.046125$ & $1.029$    \\
$51168256$ & $2.75$   & $133.2506666667$ & $206.805$  & $45.8925$  & $133.5033641667$ & $27.82375$ & $0.044$    & $0.9795$   \\ 
\boldmath{$49255227.2$} & \boldmath{$2.8158$} & \boldmath{$128.2688208333$} & \boldmath{$208.0035$} & \boldmath{$46.75425$} & \boldmath{$128.5235785833$} & \boldmath{$29.223875$} & \boldmath{$0.0466375$} & \boldmath{$1.0349375$} \\ \hline

\end{tabular}
\end{table}
\end{landscape}

\begin{landscape}
\begin{table}[]
\caption{Average Delay and Delay Jitter for different real-life Image files} \label{different_image_file}
\scriptsize
\centering
\begin{tabular}{|c|c|c|c|c|c|c|c|c|}
\hline
\parbox{1.75cm}{Image File Size (in bits)} & \parbox{1.75cm}{Average Number of Channel Deallocation by PUs} & \parbox{1.75cm}{Ideal Transmission Time (IT) (in sec)} & \parbox{1.75cm}{Initial Channel Allocation Time (ICA) (in msec)} & \parbox{1.75cm}{Channel Reallocation Time (CR) (in msec)} & \parbox{2cm}{Actual Transmission Time (AT) (in sec) (AT=IT+ICA+CR)} & \parbox{1.75cm}{Maximum Jitter (in msec)} & \parbox{1.75cm}{Mean Jitter (in msec)} & \parbox{1.75cm}{Standard Deviation In Jitter (in msec)} \\ \hline

\multicolumn{9}{|c|}{Average number of free channels ($F$) is $697$ }    \\ \hline
$12448128$ & $0.23$  & $48.6255$     & $9.235$ & $0.34625$ & $48.63508125$ & $0.325$   & $0.0009111875$ & $0.017$    \\
$12459888$ & $0.232$ & $48.6714375$  & $9.125$ & $0.385$   & $48.6809475$  & $0.34625$ & $0.0010105$    & $0.018375$ \\
$12486256$ & $0.199$ & $48.7744375$  & $9.165$ & $0.33125$ & $48.78393375$ & $0.31$    & $0.00086715$   & $0.01625$  \\
$12596912$ & $0.248$ & $49.2066875$  & $9.275$ & $0.385$   & $49.2163475$  & $0.3475$  & $0.001$        & $0.018375$ \\
$12650768$ & $0.234$ & $49.4170625$  & $9.215$ & $0.38125$ & $49.42665875$ & $0.3475$  & $0.0009851375$ & $0.01825$  \\
$12710512$ & $0.221$ & $49.6504375$  & $9.11$  & $0.35375$ & $49.65990125$ & $0.32875$ & $0.000911725$  & $0.017125$ \\
$12828824$ & $0.24$  & $50.11259375$ & $9.095$ & $0.385$   & $50.12207375$ & $0.345$   & $0.0009821375$ & $0.018$    \\
$12903144$ & $0.249$ & $50.40290625$ & $9.08$  & $0.4$     & $50.41238625$ & $0.36$    & $0.001015225$  & $0.01875$  \\
$12927384$ & $0.25$  & $50.49759375$ & $9.22$  & $0.38375$ & $50.5071975$  & $0.35625$ & $0.000971525$  & $0.018375$ \\
$12996424$ & $0.263$ & $50.76728125$ & $9.19$  & $0.43375$ & $50.776905$   & $0.395$   & $0.001092575$  & $0.0205$   \\ 
\boldmath{$12700824$} & \boldmath{$0.2366$} & \boldmath{$49.61259375$} & \boldmath{$9.171$} & \boldmath{$0.3785$} & \boldmath{$49.62214325$} & \boldmath{$0.346125$} & \boldmath{$0.0009747163$} & \boldmath{$0.0181$} \\ \hline

\multicolumn{9}{|c|}{Average number of free channels ($F$) is $510$ }    \\ \hline
$12448128$ & $0.331$ & $48.6255$     & $11.57$  & $0.7575$  & $48.6378275$  & $0.68875$ & $0.002$    & $0.036375$ \\
$12459888$ & $0.311$ & $48.6714375$  & $11.755$ & $0.63125$ & $48.68382375$ & $0.57125$ & $0.001625$ & $0.03025$  \\
$12486256$ & $0.32$  & $48.7744375$  & $11.69$  & $0.68$    & $48.7868075$  & $0.60375$ & $0.00175$  & $0.032$    \\ 
$12596912$ & $0.342$ & $49.2066875$  & $11.775$ & $0.72625$ & $49.21918875$ & $0.655$   & $0.001875$ & $0.034375$ \\ 
$12650768$ & $0.302$ & $49.4170625$  & $11.905$ & $0.66125$ & $49.42962875$ & $0.59375$ & $0.00175$  & $0.031$    \\ 
$12710512$ & $0.309$ & $49.6504375$  & $11.675$ & $0.67125$ & $49.66278375$ & $0.6125$  & $0.00175$  & $0.031875$ \\ 
$12828824$ & $0.31$  & $50.11259375$ & $11.93$  & $0.66$    & $50.12518375$ & $0.595$   & $0.001625$ & $0.030875$ \\ 
$12903144$ & $0.327$ & $50.40290625$ & $11.655$ & $0.69875$ & $50.41526$    & $0.62125$ & $0.00175$  & $0.032375$ \\ 
$12927384$ & $0.306$ & $50.49759375$ & $11.58$  & $0.68375$ & $50.5098575$  & $0.6125$  & $0.00175$  & $0.031875$ \\ 
$12996424$ & $0.313$ & $50.76728125$ & $11.75$  & $0.71375$ & $50.779745$   & $0.64625$ & $0.00175$  & $0.033375$ \\ 
\boldmath{$12700824$} & \boldmath{$0.3171$} & \boldmath{$49.61259375$} & \boldmath{$11.7285$} & \boldmath{$0.688375$} & \boldmath{$49.625010625$} & \boldmath{$0.62$} & \boldmath{$0.0017625$} & \boldmath{$0.0324375$} \\ \hline

\multicolumn{9}{|c|}{Average number of free channels ($F$) is $285$ }    \\ \hline
$12448128$ & $0.501$ & $48.6255$     & $19.315$ & $1.72875$ & $48.64654375$ & $1.51375$ & $0.0045$   & $0.0805$   \\
$12459888$ & $0.464$ & $48.6714375$  & $19.395$ & $1.565$   & $48.6923975$  & $1.335$   & $0.004125$ & $0.071625$ \\
$12486256$ & $0.44$  & $48.7744375$  & $19.39$  & $1.6675$  & $48.795495$   & $1.48$    & $0.004375$ & $0.078375$ \\
$12596912$ & $0.497$ & $49.2066875$  & $18.76$  & $1.705$   & $49.2271525$  & $1.4875$  & $0.004375$ & $0.078625$ \\
$12650768$ & $0.444$ & $49.4170625$  & $19.865$ & $1.6275$  & $49.438555$   & $1.3925$  & $0.00425$  & $0.073875$ \\
$12710512$ & $0.469$ & $49.6504375$  & $19.665$ & $1.7825$  & $49.671885$   & $1.55375$ & $0.004625$ & $0.0815$   \\
$12828824$ & $0.511$ & $50.11259375$ & $19.61$  & $1.91625$ & $50.13412$    & $1.695$   & $0.004875$ & $0.08825$  \\ 
$12903144$ & $0.524$ & $50.40290625$ & $19.725$ & $1.96875$ & $50.4246$     & $1.69125$ & $0.005$    & $0.0885$   \\
$12927384$ & $0.471$ & $50.49759375$ & $19.645$ & $1.7075$  & $50.51894625$ & $1.46125$ & $0.004375$ & $0.0765$   \\
$12996424$ & $0.524$ & $50.76728125$ & $19.48$  & $1.83875$ & $50.7886$     & $1.55125$ & $0.004625$ & $0.081375$ \\ 
\boldmath{$12700824$} & \boldmath{$0.4845$} & \boldmath{$49.61259375$} & \boldmath{$19.485$} & \boldmath{$1.75075$} & \boldmath{$49.6338295$} & \boldmath{$1.516125$} & \boldmath{$0.0045125$} & \boldmath{$0.0799125$} \\ \hline

 \multicolumn{9}{|c|}{Average number of free channels ($F$) is $39$ }    \\ \hline
$12448128$ & $1.455$ & $48.6255$     & $137.89$  & $27.355$   & $48.790745$   & $19.87375$ & $0.072$    & $1.10475$  \\
$12459888$ & $1.429$ & $48.6714375$  & $131.225$ & $27.9175$  & $48.83058$    & $21.2225$  & $0.07325$  & $1.158875$ \\
$12486256$ & $1.428$ & $48.7744375$  & $135.71$  & $27.91375$ & $48.93806125$ & $21.02$    & $0.073125$ & $1.1525$   \\
$12596912$ & $1.463$ & $49.2066875$  & $138.17$  & $28.785$   & $49.3736425$  & $20.625$   & $0.07475$  & $1.146125$ \\
$12650768$ & $1.465$ & $49.4170625$  & $133.615$ & $27.2375$  & $49.577915$   & $20.34$    & $0.070375$ & $1.112375$ \\
$12710512$ & $1.4$   & $49.6504375$  & $138.69$  & $26.57125$ & $49.81569875$ & $20.06125$ & $0.0685$   & $1.091125$ \\
$12828824$ & $1.396$ & $50.11259375$ & $136.1$   & $25.3825$  & $50.27407625$ & $18.91625$ & $0.06475$  & $1.02425$  \\
$12903144$ & $1.419$ & $50.40290625$ & $134.345$ & $26.0275$  & $50.56327875$ & $18.91625$ & $0.066$    & $1.033$    \\
$12927384$ & $1.501$ & $50.49759375$ & $136.59$  & $27.545$   & $50.66172875$ & $19.88125$ & $0.06975$  & $1.085375$ \\
$12996424$ & $1.425$ & $50.76728125$ & $136.87$  & $27.82375$ & $50.931975$   & $20.9325$  & $0.070125$ & $1.123125$ \\ 
\boldmath{$12700824$} & \boldmath{$1.4381$} & \boldmath{$49.61259375$} & \boldmath{$135.9205$} & \boldmath{$27.255875$} & \boldmath{$49.775770125$} & \boldmath{$20.178875$} & \boldmath{$0.0702625$} & \boldmath{$1.10315$} \\ \hline

\end{tabular}
\end{table}
\end{landscape}

\begin{landscape}
\begin{table}[]
\caption{Average Delay and Delay Jitter for different real-life Text files} \label{different_data_file}
\scriptsize
\centering
\begin{tabular}{|c|c|c|c|c|c|c|c|c|}
\hline
\parbox{1.75cm}{Data File Size (in bits)} & \parbox{1.75cm}{Average Number of Channel Deallocation by PUs} & \parbox{1.75cm}{Ideal Transmission Time (IT) (in sec)} & \parbox{1.75cm}{Initial Channel Allocation Time (ICA) (in msec)} & \parbox{1.75cm}{Channel Reallocation Time (CR) (in msec)} & \parbox{2cm}{Actual Transmission Time (AT) (in sec) (AT=IT+ICA+CR)} & \parbox{1.75cm}{Maximum Jitter (in msec)} & \parbox{1.75cm}{Mean Jitter (in msec)} & \parbox{1.75cm}{Standard Deviation In Jitter (in msec)} \\ \hline

\multicolumn{9}{|c|}{Average number of free channels ($F$) is $697$ }    \\ \hline
$25848$   & $0$     & $0.2019375$  & $3.9925$ & $0$       & $0.20593$     & $0$       & $0$            & $0$        \\
$29008$   & $0$     & $0.226625$   & $4.1225$ & $0$       & $0.2307475$   & $0$       & $0$            & $0$        \\
$30752$   & $0$     & $0.24025$    & $4.035$  & $0$       & $0.244285$    & $0$       & $0$            & $0$        \\
$112920$  & $0.004$ & $0.8821875$  & $4.11$   & $0.00625$ & $0.88630375$  & $0.00625$ & $0.0008928625$ & $0.002375$ \\
$322448$  & $0.009$ & $2.519125$   & $4.0375$ & $0.015$   & $2.5231775$   & $0.015$   & $0.00075$      & $0.003375$ \\
$453664$  & $0.014$ & $3.54425$    & $3.915$  & $0.0275$  & $3.5481925$   & $0.0275$  & $0.0009821375$ & $0.00525$  \\
$564880$  & $0.013$ & $4.413125$   & $3.9925$ & $0.02625$ & $4.41714375$  & $0.02625$ & $0.00075$      & $0.004375$ \\
$703728$  & $0.018$ & $5.497875$   & $4.0775$ & $0.0275$  & $5.50198$     & $0.0275$  & $0.0006395375$ & $0.00425$  \\
$1284416$ & $0.029$ & $10.0345$    & $3.9675$ & $0.05125$ & $10.03851875$ & $0.05125$ & $0.0006487375$ & $0.00575$  \\
$3098904$ & $0.06$  & $24.2101875$ & $3.9575$ & $0.10625$ & $24.21425125$ & $0.105$   & $0.0005592125$ & $0.007625$ \\ 
\boldmath{$662656.8$} & \boldmath{$0.0147$} & \boldmath{$5.17700625$} & \boldmath{$4.02075$} & \boldmath{$0.026$} & \boldmath{$5.181053$} & \boldmath{$0.025875$} & \boldmath{$0.0005222488$} & \boldmath{$0.0033$} \\ \hline

\multicolumn{9}{|c|}{Average number of free channels ($F$) is $510$ }    \\ \hline
$25848$   & $0.003$ & $0.2019375$  & $5.415$  & $0.0075$  & $0.20736$    & $0.0075$  & $0.00375$      & $0.00525$  \\ 
$29008$   & $0$     & $0.226625$   & $5.565$  & $0$       & $0.23219$    & $0$       & $0$            & $0$        \\ 
$30752$   & $0$     & $0.24025$    & $5.495$  & $0$       & $0.245745$   & $0$       & $0$            & $0$        \\ 
$112920$  & $0.003$ & $0.8821875$  & $5.405$  & $0.00625$ & $0.88759875$ & $0.00625$ & $0.0008928625$ & $0.002375$ \\ 
$322448$  & $0.008$ & $2.519125$   & $5.505$  & $0.02$    & $2.52465$    & $0.02$    & $0.001$        & $0.0045$   \\ 
$453664$  & $0.016$ & $3.54425$    & $5.4275$ & $0.03625$ & $3.54971375$ & $0.03625$ & $0.00125$      & $0.006875$ \\ 
$564880$  & $0.012$ & $4.413125$   & $5.6$    & $0.02375$ & $4.41874875$ & $0.02375$ & $0.000678575$  & $0.004$    \\ 
$703728$  & $0.03$  & $5.497875$   & $5.4475$ & $0.06375$ & $5.50338625$ & $0.06$    & $0.0015$       & $0.009375$ \\ 
$1284416$ & $0.04$  & $10.0345$    & $5.5$    & $0.0975$  & $10.0400975$ & $0.09375$ & $0.001234175$  & $0.01075$  \\ 
$3098904$ & $0.076$ & $24.2101875$ & $5.3725$ & $0.1825$  & $24.2157425$ & $0.175$   & $0.000960525$  & $0.012875$ \\ 
\boldmath{$662656.8$} & \boldmath{$0.0188$} & \boldmath{$5.17700625$} & \boldmath{$5.47325$} & \boldmath{$0.04375$} & \boldmath{$5.18252325$} & \boldmath{$0.04225$} & \boldmath{$0.0011266138$} & \boldmath{$0.0056$} \\ \hline

\multicolumn{9}{|c|}{Average number of free channels ($F$) is $285$ }    \\ \hline
$25848$   & $0.002$ & $0.2019375$  & $9.1725$ & $0.00375$ & $0.21111375$  & $0.00375$ & $0.001875$ & $0.002625$ \\
$29008$   & $0.001$ & $0.226625$   & $9.54$   & $0.00375$ & $0.23616875$  & $0.00375$ & $0.001875$ & $0.002625$ \\
$30752$   & $0.001$ & $0.24025$    & $9.6875$ & $0.0025$  & $0.24994$     & $0.0025$  & $0.00125$  & $0.00175$  \\
$112920$  & $0.006$ & $0.8821875$  & $9.2875$ & $0.02125$ & $0.89149625$  & $0.02125$ & $0.003$    & $0.008$    \\
$322448$  & $0.018$ & $2.519125$   & $9.3875$ & $0.0875$  & $2.5286$      & $0.0875$  & $0.004375$ & $0.019625$ \\
$453664$  & $0.028$ & $3.54425$    & $9.2975$ & $0.08875$ & $3.55363625$  & $0.0875$  & $0.003125$ & $0.0165$   \\
$564880$  & $0.036$ & $4.413125$   & $9.7025$ & $0.17125$ & $4.42299875$  & $0.17125$ & $0.004875$ & $0.029$    \\
$703728$  & $0.038$ & $5.497875$   & $9.495$  & $0.11125$ & $5.50748125$  & $0.11125$ & $0.002625$ & $0.017$    \\
$1284416$ & $0.057$ & $10.0345$    & $9.5725$ & $0.28125$ & $10.04435375$ & $0.2775$  & $0.003625$ & $0.0315$   \\
$3098904$ & $0.135$ & $24.2101875$ & $9.2675$ & $0.54625$ & $24.22000125$ & $0.5175$  & $0.002875$ & $0.038125$ \\ 
\boldmath{$662656.8$} & \boldmath{$0.0322$} & \boldmath{$5.17700625$} & \boldmath{$9.441$} & \boldmath{$0.13175$} & \boldmath{$5.186579$} & \boldmath{$0.128375$} & \boldmath{$0.00295$} & \boldmath{$0.016675$} \\ \hline

\multicolumn{9}{|c|}{Average number of free channels ($F$) is $39$ }    \\ \hline
$25848$   & $0.008$ & $0.2019375$  & $65.48$   & $0.22875$ & $0.26764625$ & $0.22875$ & $0.114375$ & $0.16175$  \\
$29008$   & $0.011$ & $0.226625$   & $66.48$   & $0.3425$  & $0.2934475$  & $0.3425$  & $0.17125$  & $0.242125$ \\
$30752$   & $0.01$  & $0.24025$    & $63.5325$ & $0.37625$ & $0.30415875$ & $0.37625$ & $0.188125$ & $0.266$    \\
$112920$  & $0.044$ & $0.8821875$  & $64.85$   & $1.05375$ & $0.94809125$ & $1.05375$ & $0.1505$   & $0.39825$  \\
$322448$  & $0.13$  & $2.519125$   & $66.66$   & $4.40625$ & $2.59019125$ & $4.33375$ & $0.220375$ & $0.973$    \\
$453664$  & $0.162$ & $3.54425$    & $65.68$   & $5.44375$ & $3.61537375$ & $5.19125$ & $0.194375$ & $0.994125$ \\
$564880$  & $0.204$ & $4.413125$   & $66.3075$ & $6.0925$  & $4.485525$   & $5.8725$  & $0.174125$ & $1.001125$ \\
$703728$  & $0.215$ & $5.497875$   & $66.3075$ & $6.04$    & $5.5702225$  & $5.74125$ & $0.1405$   & $0.884625$ \\
$1284416$ & $0.358$ & $10.0345$    & $66.82$   & $9.1375$  & $10.1104575$ & $8.3625$  & $0.115625$ & $0.962125$ \\
$3098904$ & $0.508$ & $24.2101875$ & $66.67$   & $10.94$   & $24.2877975$ & $9.53625$ & $0.057625$ & $0.716625$ \\ 
\boldmath{$662656.8$} & \boldmath{$0.165$} & \boldmath{$5.17700625$} & \boldmath{$65.87875$} & \boldmath{$4.406125$} & \boldmath{$5.247291125$} & \boldmath{$4.103875$} & \boldmath{$0.1526875$} & \boldmath{$0.659975$} \\ \hline

\end{tabular}
\end{table}
\end{landscape}

\bibliographystyle{plain}
\bibliography{reff}

\end{document}